\documentclass[%
 aip,
 amsmath,amssymb,
 reprint,%
]{revtex4-1}

\usepackage{graphicx}% Include figure files
\usepackage{dcolumn}% Align table columns on decimal point
\usepackage{bm}% bold math
\usepackage{eucal}
\usepackage[utf8]{inputenc}
\usepackage[T1]{fontenc}
\usepackage{dsfont}
\usepackage{mathptmx}
\usepackage{etoolbox}
\usepackage{xcolor}
\usepackage{amsmath}
\usepackage{epstopdf}
\usepackage{graphicx}
\usepackage{epstopdf}
\usepackage[colorlinks=true, linkcolor=blue, citecolor=blue, urlcolor=blue]{hyperref}

\DeclareMathOperator{\sech}{sech}

\makeatletter
\def\@email#1#2{%
 \endgroup
 \patchcmd{\titleblock@produce}
  {\frontmatter@RRAPformat}
  {\frontmatter@RRAPformat{\produce@RRAP{*#1\href{mailto:#2}{#2}}}\frontmatter@RRAPformat}
  {}{}
}%
\makeatother
\begin{document}

\preprint{AIP/123-QED}

\title[]{Coherent Control of Population and Quantum Coherence in Superconducting Circuits }
% Force line breaks with \\
\author{Madan Mohan Mahana}
\affiliation{Department of Physics, Indian Institute of Technology Guwahati, Guwahati 781039, Assam, India}
\author{Gunjan Yadav}
\affiliation{Department of Physics, Indian Institute of Technology Guwahati, Guwahati 781039, Assam, India}
\author{Tarak Nath Dey}
\email[Corresponding author: ]{tarak.dey@iitg.ac.in}
\affiliation{Department of Physics, Indian Institute of Technology Guwahati, Guwahati 781039, Assam, India}

\date{\today}% It is always \today, today,
             %  but any date may be explicitly specified

\begin{abstract}
Quantum mechanics, with its counterintuitive principles and probabilistic nature, has long been confined to the microscopic realm of atoms and photons. Yet, recent breakthroughs have pushed the boundaries of quantum behavior into the macroscopic world—where objects are visible to the naked eye and governed by classical physics. This review article traces the extraordinary progress toward achieving coherent control of population distributions among multiple quantum levels, as well as manipulation of absorption and refractive index, in such large-scale quantum systems—a feat once considered beyond reach.
\end{abstract}

\maketitle

\section{\label{sec:level1} Introduction}
The study of quantum‑mechanical phenomena at macroscopic scales \cite{PhysRevLett.53.1260,PhysRevLett.55.1908,PhysRevLett.55.1543} opens new possibilities for building quantum computers \cite{RN17322,Kandala2017HWEVQE,Barends2014SurfaceCode,Mooij2005RoadToQC}. Superconducting circuit quantum electrodynamics (cQED), based on Josephson junctions \cite{JOSEPHSON1962251,Josephson01101965,Leggett1980MacroscopicQM,LeggettGarg1985MacroscopicRealism,Leggett1999MQCSignificance}, has emerged as a leading platform for solid‑state quantum computation and quantum information processing \cite{Nielsen_Chuang_2010,RN20914,RN20916,RevModPhys.73.357}. Quantum coherence \cite{RN10745,Arimondo1996CPT,PhysRevLett.56.1811,PhysRevLett.66.2593}, quantum entanglement \cite{RevModPhys.71.S288,RN20404,Berkley2003EntangledQubits}, and quantum parallelism\cite{Shor_1997,Collins1997QuantumMagic,Grover1997SearchAlgorithm,Brassard1997QuantumPhoneBook} provide quantum computers with abilities that surpass those of classical computers for specific tasks, such as integer factorization \cite{365700,Nielsen_Chuang_2010}, fast quantum search \cite{Grover1997SearchAlgorithm,Brassard1997QuantumPhoneBook,Collins1997QuantumMagic} and the efficient simulation of complex quantum many‑body systems \cite{Fauseweh2024QuantumManyBody}. The idea of the quantum simulator was first introduced by Richard Feynman \cite{RN20917}, and two years later David Deutsch demonstrated the solution of a toy problem using a quantum algorithm \cite{Deutsch_1983}. \\
\indent The qubit serves as the fundamental building block of a quantum computer. In superconducting cQED systems, the two lowest energy levels of an engineered quantum circuit can be utilized to form a quantum bit \cite{RN20918,RN20919,YouNori2005,SchoelkopfGirvin2008, 10.1093/acprof:oso/9780199681181.003.0003}. These artificial two‑level systems exhibit quantum interference \cite{Friedman2000MacroscopicSuperposition,vanderWal2000PersistentCurrentSuperposition,Berkley2003EntangledQubits} and coherent dynamics \cite{doi:10.1126/science.285.5430.1036, Bocko1997ProspectsQC, Ioffe1999QuietQubit, PhysRevB.60.15398} that can be precisely controlled through macroscopic electrical parameters such as voltages or currents. In the simplest realization, the qubit states correspond to distinct charge configurations, coupled through the tunneling of Cooper pairs across a Josephson junction \cite{Girvin2009CircuitQED}.\\
\indent Among the various superconducting qubit architectures, the transmon qubit \cite{PhysRevA.76.042319} based on the Cooper-Pair Box (CPB) \cite{Cooper1956BoundPairs} is a particularly promising candidate for realizing a robust and controllable qubit \cite{RN20914,RN20916,RevModPhys.73.357}. The transmon qubit is commonly used in state-of-the-art Noisy Intermediate-Scale Quantum (NISQ) processors \cite{Acharya2025, Kim2023}, owing to its reduced sensitivity to charge noise \cite{PhysRevB.77.180502}, higher coherence time \cite{Bland2025}, scalability \cite{Kosen_2022}, and robustness \cite{doi:10.1126/science.abb2823}. Importantly, these flexibilities enable the cQED system to reach the strong‑coupling regime of light–matter interaction \cite{Wallraff2004CircuitQED,Majer2007CavityBus,RevModPhys.93.025005} far more readily than is possible with traditional atomic \cite{Tey2008StrongInteraction} , ions \cite{DeVoe1996Superradiance}, and quantum dots platforms \cite{Vamivakas2007StrongExtinction,Muller2007ResonanceFluorescence}. Fluxonium qubit \cite{doi:10.1126/science.1175552} with coherence time on par or exceeding that of transmon, and large anharmonicity is another emerging platform for next-generation quantum processors \cite{PRXQuantum.3.037001, PhysRevApplied.21.024015}.\\
\indent In quantum optics, several counterintuitive phenomena—such as electromagnetically induced transparency (EIT)\cite{PhysRevLett.64.1107,PhysRevLett.66.2593,RevModPhys.77.633,PhysRevLett.56.1811}, Autler–Townes splitting (ATS)\cite{AutlerTownes1955,Wu1975PRL,Schabert1975ApplPhys}, coherent population trapping (CPT)\cite{RN10745,Arimondo1996CPT}, coherent population oscillation (CPO)\cite{Wu1977PRL,Gruneisen1988JOSAB,Davis1996PRA,AgarwalDey2009SlowLight}, and stimulated Raman adiabatic passage (STIRAP)\cite{Gaubatz1988CPL, RevModPhys.89.015006}—play pivotal roles in enabling precise control over the optical properties of a medium. These coherence‑driven effects require atomic systems with long-lived coherence between relevant energy levels. In particular, a three‑level configuration containing two metastable lower states provides the necessary long coherence times to observe and exploit such phenomena. Consequently, three‑level systems that exhibit strong anharmonicity and a minimal number of rapidly decaying states are especially well suited for realizing these quantum‑optical effects. Historically, most demonstrations of these effects were performed using natural atoms, often placed inside optical resonators formed by high‑reflectivity mirrors—an arrangement that laid the foundation for cavity quantum electrodynamics (CQED)\cite{Miller2005TrappedAtomsCQED,Walther2006CQED}. However, the landscape of quantum optics has expanded dramatically with the advent of superconducting quantum circuits (SQCs). The striking features of superconducting cQED devices is that they can be fabricated and integrated at macroscopic scales, enabling straightforward scalability and supporting rapid quantum‑state manipulation. Importantly, superconducting circuits can be designed to realize highly coherent three-level structures, making them powerful and versatile platforms for exploring quantum interference phenomena.  Indeed, three‑level superconducting artificial atoms have already been used to experimentally demonstrate EIT \cite{PhysRevLett.120.083602,Murali2004EITdecoherence,Dutton2006EITsuperconducting}, ATS \cite{PhysRevLett.102.243602}, and CPT \cite{Kelly2010}, showcasing their potential as a leading architecture for next‑generation quantum‑optics and quantum‑information technologies.\\
\indent Beyond superconducting cQED platforms, several other physical systems are capable of supporting quantum superposition and coherent control of qubits for building quantum computers. These include quantum dots \cite{Imamoglu1999Quant-5261,Petta2005Coherent,Englund2005Controlling,Hanson2007Spins}, spin defects \cite{Hanson2006RoomTemp,Dutt2007QuantumRegister, 93lt-gj9k}, trapped ions\cite{CiracZoller1995ColdIons,Leibfried2003SingleIons,Blatt2008Entangled,Haeffner2008QCTrappedIons,Blatt2012QuantumSimulations}, ultracold atoms \cite{Jaksch2005HubbardToolbox,Lewenstein2007UltracoldLattices,Bloch2008ManyBodyUltracold,Gross2017QuantumSimulations}, and polarized photons\cite{Knill2001LinearOpticsQC,Pittman2001Probabilistic,Franson2002HighFidelity,Pittman2003CNOT}. In each of these architectures, microscopic quantum systems are used to encode quantum information, and each approach carries its own advantages and limitations \cite{RN17322}.

This review article provides a foundational introduction to quantum optics in superconducting quantum circuits (SQCs). In Sec. \ref{Sec:II}, we perform the quantization of the electromagnetic (EM) field in vacuum, an LC circuit, and a one-dimensional transmission line. We then briefly introduce decoherence in an open quantum system (OQS) and the Lindblad master equation, followed by a discussion of coherence in a two-level quantum system. In Sec. \ref{Sec:III}, we first provide a basic introduction to superconductivity and the Josephson junction. Then, we move on to introduce the three types of superconducting qubits, namely charge qubit, flux qubit, and phase qubit. In Sec. \ref{Sec:IV}, the implementation of the Jaynes-Cummings model in a coupled transmon-LC oscillator system is discussed, which is further utilized for engineering a three-level $\Delta$ system in the doubly-dressed polatiton states basis by driving the transmon qubit. With all the transitions in the $\Delta$ system electric-dipole-allowed, we explore the realization of quantum phenomena based on coherent control of quantum coherence, such as EIT and ATS. We also discuss the coherent control of population transfer via adiabatic STIRAP and shortcut-to-adiabaticity (STA) aided superadiabatic STIRAP (saSTIRAP) in the dressed-state-engineered $\Delta$ system in circuit QED. Finally in Sec. \ref{Sec:V}, we summarize our discussions with a conclusion. \\
\section{Quantization and Dissipation}\label{Sec:II}
Modern quantum technologies rely on efficient light–matter interaction\cite{weiner2008light} and the precise control and manipulation of EM fields and atomic systems \cite{PhysRevA.84.032116}. To understand this interaction properly, we need two important ideas: quantization and dissipation. Quantization means that physical quantities such as energy are not continuous, but come in discrete units called quanta \cite{biswas2023quantization}. Applying quantization to EM waves\cite{landi2014quantisation} and electrical circuits\cite{10.1063/1.4937246}, we can describe light in terms of photons and treat circuits as quantum systems, which is essential for studying atoms, cavities\cite{PhysRevA.67.013805}, waveguides \cite{halir2015waveguide,dkbz-thfh}, and superconducting devices\cite{ruggiero2013superconducting}.
However, real physical systems are never perfectly isolated. They always interact with their surroundings, which leads to energy loss and loss of coherence \cite{menskii2003dissipation}. This effect is known as dissipation \cite{celeghini1992quantum}. To accurately describe realistic systems, especially in experiments, we must include these dissipative effects in our theory. Therefore, in this section, we introduce both quantization and dissipation to build a complete and practical framework for studying quantum systems.
\subsection{Quantization of Electromagnetic Fields}
In classical electrodynamics, Maxwell’s equations successfully describes classical EM phenomena such as wave propagation, interference, and radiation \cite{jackson2012classical}. However, the classical picture treats the EM field as a continuous classical quantity, which fails to explain key quantum effects such as spontaneous emission\cite{weisskopf1930berechnung}, vacuum fluctuations\cite{PhysRev.72.241}, and the Casimir effect \cite{casimir1948attraction}. These phenomena demonstrate that even in the absence of any sources of radiation, the EM field exhibits irreducible quantum fluctuations \cite{loudon1974quantum, scully1997quantum}. Thus, a full understanding of these phenomena requires quantization of the EM field.
An ideal platform for quantization of EM field are confined structures like cavities and waveguides\cite{PhysRevA.67.013805}. Boundary conditions in these systems restrict the field into discrete normal modes, each mode behaving as an independent quantum harmonic oscillator. This mode based description helps us to understand field quantization in free space and serves as the starting point for quantizing electrical circuits. 
To develop this idea more clearly, let us first consider a classical EM in free space in the absence of sources (charges and currents)\cite{mandel1996optical}. The field satisfies the homogeneous Maxwell equations (in SI units):
\begin{subequations}
\begin{align}
\nabla \cdot \mathbf{E}(\mathbf{r},t) &= 0, \label{eq:gaussE} \\[4pt]
\nabla \cdot \mathbf{B}(\mathbf{r},t) &= 0, \label{eq:gaussB} \\[4pt]
\nabla \times \mathbf{E}(\mathbf{r},t) &= -\frac{\partial \mathbf{B}(\mathbf{r},t)}{\partial t}, \label{eq:faraday} \\[4pt]
\nabla \times \mathbf{B}(\mathbf{r},t) &= \frac{1}{c^2}\frac{\partial \mathbf{E}(\mathbf{r},t)}{\partial t}. \label{eq:ampere}
\end{align}
\end{subequations}
Here, $\mathbf{E}(\mathbf{r},t)$ and $\mathbf{B}(\mathbf{r},t)$ are electric and magnetic field vectors at the space-time point $(\mathbf{r},t)$. These fields can be expressed in terms of transverse vector potential $\mathbf{A}(\mathbf{r},t)$ in Coulomb gauge, which satisfies the homogeneous wave equation,
\begin{equation}
\nabla^2 \mathbf{A}(\mathbf{r},t) - \frac{1}{c^2}\frac{\partial^2 \mathbf{A}(\mathbf{r},t)}{\partial t^2} = 0,
\end{equation}
and satisfies the divergence condition, $\nabla \cdot \mathbf{A}(\mathbf{r},t) = 0$. In terms of $\mathbf{A}(\mathbf{r},t)$, the electric and magnetic fields $\mathbf{E}(\mathbf{r},t)$ and $\mathbf{B}(\mathbf{r},t)$ are given by
\begin{subequations}
\begin{align}
\mathbf{E}(\mathbf{r},t) &= -\frac{\partial \mathbf{A}(\mathbf{r},t)}{\partial t}, 
\label{Eq: 2a}\\[4pt]
\mathbf{B}(\mathbf{r},t) &= \nabla \times \mathbf{A}(\mathbf{r},t),
\label{Eq: 2b}
\end{align}
\end{subequations} 
%%together with the divergence condition
%%\begin{equation}
%%\nabla \cdot \mathbf{A}(\mathbf{r},t) = 0.
%%\end{equation}
%%The electric $\mathbf{E}(\mathbf{r},t)$ %%%and magnetic field $\mathbf{B}%%(\mathbf{r},t)$ in terms of $\mathbf{A}%%(\mathbf{r},t)$ are given by
%%\begin{subequations}
%%\begin{align}
%%\mathbf{E}(\mathbf{r},t) &= -%%\frac{\partial \mathbf{A}(\mathbf{r},t)}%%{\partial t}, \\[4pt]
%%\mathbf{B}(\mathbf{r},t) &= \nabla \times %%\mathbf{A}(\mathbf{r},t).
%%\end{align}
%%\end{subequations}
To quantize the EM, we start by expressing the vector potential $\mathbf{A}(\mathbf{r},t)$ in terms of its discrete Fourier components. This is done under the assumption that the field is confined within a large cubic box of side length $L$, with periodic boundary conditions. This setup simplifies the mathematics and allows us to treat the field as a sum over discrete modes.
%To get the Hamiltonian equations of motion, it is useful to make a Fourier decomposition of $\mathbf{A}(\mathbf{r},t)$ with respect to its spatial variable:
\begin{equation}
\mathbf{A}(\mathbf{r},t) = \frac{1}{ \epsilon_0 ^{1/2} L^{3/2}}\sum_{\mathbf{k}} \boldsymbol{\mathcal{A}}_{\mathbf{k}}(t)e^{i\mathbf{k}\cdot\mathbf{r}},
\end{equation}
where the components of $\mathbf{k}$ are
\begin{equation}
    k_i = \frac{2\pi n_i}{L},\quad i\in{x,y,z} \quad n_i = 0, \pm 1, \pm 2, \dots, 
\end{equation}
forming a discrete set. The sum $\sum_{\mathbf{k}}$ is understood to be sum over all integers $(n_x, n_y, n_z)$. The factor $\epsilon_0 ^{1/2} L^{3/2}$ ensures correct normalization, where $\epsilon_0$ is the vacuum dielectric constant.
Using the divergence condition, one can get
\begin{equation}
\mathbf{k} \cdot \boldsymbol{\mathcal{A}}_{\mathbf{k}}(t) = 0.
\label{Eq: 6}
\end{equation}
In addition, the reality of $\mathbf{A}(\mathbf{r},t)$ leads to
\begin{equation}
\boldsymbol{\mathcal{A}}_{-\mathbf{k}}(t) = \boldsymbol{\mathcal{A}}_{\mathbf{k}}^*(t).
\label{Eq: 7}
\end{equation}
Since $\mathbf{A}(\mathbf{r},t)$ satisfies the wave equation, it follows
\begin{equation}
\left( \frac{\partial^2}{\partial t^2} + \omega_\mathbf{k}^2 \right)\boldsymbol{\mathcal{A}}_{\mathbf{k}}(t) = 0,
\end{equation}
where $\omega_\mathbf{k}$ is the angular frequency and defined as $\omega_k = c|\mathbf{k}|$. The general solution of this equation, obeying the  Eq. \eqref{Eq: 7} are
\begin{equation}
\boldsymbol{\mathcal{A}}_{\mathbf{k}}(t) = \mathbf{c}_{\mathbf{k}} e^{-i\omega_k t} + \mathbf{c}_{-\mathbf{k}}^* e^{i\omega_k t}.
\label{Eq: 9}
\end{equation}
The vector $\mathbf{c}_{\mathbf{k}}$ can be decomposed into two orthogonal polarization components so that Eq. \eqref{Eq: 6} is satisfied automatically. We can do it most easily by selecting a pair of orthogonal real base vectors $\mathbf{\epsilon}_{\mathbf{k}1}$, $\mathbf{\epsilon}_{\mathbf{k}2}$
\begin{equation}
\mathbf{c}_{\mathbf{k}} = \sum_{s=1}^2 c_{\mathbf{k}s} \boldsymbol{\epsilon}_{\mathbf{k}s},
\end{equation}
that obeys the condition
\begin{equation}
\mathbf{k}\cdot\boldsymbol{\epsilon}_{\mathbf{k}s}= 0, \quad
\boldsymbol{\epsilon}^*_{\mathbf{k}s}\cdot\boldsymbol{\epsilon}_{\mathbf{k}s'} = \delta_{ss'}, \quad
\boldsymbol{\epsilon}_{\mathbf{k}1}\times\boldsymbol{\epsilon}_{\mathbf{k}2} = \frac{\mathbf{k}}{|\mathbf{k}|}.
\end{equation}
which signifies transversality, orthogonality, and right-handedness, respectively. Substituting these into Eq. \eqref{Eq: 9}, lead to a formal solution of vector potential $\mathbf{A}(\mathbf{r},t)$ as
\begin{equation}
\mathbf{A}(\mathbf{r},t) = \frac{1}{\sqrt{\epsilon_0 L^3}}\sum_{\mathbf{k}s} \left[u_{\mathbf{k}s}(t)\boldsymbol{\epsilon}_{\mathbf{k}s}e^{i\mathbf{k}\cdot\mathbf{r}} + u_{\mathbf{k}s}^*(t)\boldsymbol{\epsilon}_{\mathbf{k}s}^*e^{-i\mathbf{k}\cdot\mathbf{r}}\right],
\end{equation}
where $u_{\mathbf{k}s}(t) = c_{\mathbf{k}s} e^{-i\omega_k t}$. The expansion of $\mathbf{A}(\mathbf{r},t)$ in terms of the fundamental vector mode function $\boldsymbol{\epsilon}_{\mathbf{k}s}e^{\mathbf{k}.r}$, with complex amplitudes $u_{\mathbf{k}s}(t)$. Each mode is labelled by vector $\mathbf{k}$ and a polarization index $s$, and the corresponding mode function evidently satisfies the Helmholtz equation
\begin{equation}
    (\nabla^2 + \mathbf{k}^2)\boldsymbol{\epsilon}_{\mathbf{k}s} e^{i \mathbf{k}.r} = 0.
\end{equation}
We can immediately make use of the electric and magnetic fields expression as provided in Eq. \eqref{Eq: 2a} and \eqref{Eq: 2b}, and rewrite as
\begin{subequations}
\begin{align}
\mathbf{E}(\mathbf{r},t) &= \frac{i}{\sqrt{\epsilon_0 L^3}}\sum_{\mathbf{k}s} \omega_\mathbf{k}\left[u_{\mathbf{k}s}(t)\boldsymbol{\epsilon}_{\mathbf{k}s}e^{i\mathbf{k}\cdot\mathbf{r}} - \text{c.c.}\right], \\[4pt]
\mathbf{B}(\mathbf{r},t) &= \frac{i}{\sqrt{\epsilon_0 L^3}}\sum_{\mathbf{k}s} \left[u_{\mathbf{k}s}(t)(\mathbf{k}\times\boldsymbol{\epsilon}_{\mathbf{k}s})e^{i\mathbf{k}\cdot\mathbf{r}} - \text{c.c.}\right].
\end{align}
\end{subequations}
and the total energy of the field is given by
\begin{equation}
H = \frac{1}{2}\int_{L^3}\left[\epsilon_0 \mathbf{E}^2(\mathbf{r},t) + \frac{1}{\mu_0}\mathbf{B}^2(\mathbf{r},t)\right] d^3r.
\end{equation}
The integration extends over the space contained within the box of volume $L^3$. On substituting the $\mathbf{E}(r,t)$ and $\mathbf{B}(r,t)$ and performing the integration over space with the help of relations:
$\int_{L^3} e^{i(\mathbf{k}-\mathbf{k}').r} d^3r = L^3 \delta^3_{\mathbf{kk}'},$ and $(\mathbf{k} \times \mathbf{\epsilon}_{\mathbf{k}s}^*). (\mathbf{k} \times \mathbf{\epsilon}_{\mathbf{k}s'}) = k^2 \mathbf{\epsilon}^*_{\mathbf{k}s}.\mathbf{\epsilon}_{\mathbf{k}s'} = k^2 \delta_{ss'}$, one can retrieve
\begin{equation}
H = 2\sum_{\mathbf{k}s} \omega_\mathbf{k}^2 |u_{\mathbf{k}s}(t)|^2.
\end{equation}
which expresses the energy as a sum over modes. For field quantization, let us introduce a pair of real canonical variables,
\begin{subequations}
\begin{align}
q_{\mathbf{k}s}(t) & = [u_{\mathbf{k}s}(t) + u_{\mathbf{k}s}^*(t)], \\
p_{\mathbf{k}s}(t) & = -i\omega_k\left[u_{\mathbf{k}s}(t) - u_{\mathbf{k}s}^*(t)\right],
\end{align}
\end{subequations}
In terms of $q_{\mathbf{k}s}(t)$ and $p_{\mathbf{k}s}(t)$ the expression for energy becomes
\begin{equation}
H = \frac{1}{2}\sum_{\mathbf{k}s} \left[p_{\mathbf{k}s}^2(t) + \omega_\mathbf{k}^2 q_{\mathbf{k}s}^2(t)\right],
\end{equation}
corresponding to a set of independent harmonic oscillators, one for each $\mathbf{k}$ and $s$ mode of the EM. The state of the classical radiation field is specified by the set of all $q_{\mathbf{k}s}(t)$, $p_{\mathbf{k}s}(t)$. The set is infinite, but since we are dealing with a finite volume and a discrete set of modes, it is countably infinite.
The field operators in terms of canonical variables are defined as
\begin{subequations}
\begin{align}
\mathbf{A}(\mathbf{r},t)
&= \frac{1}{2 \sqrt{\epsilon_0 L^3}}\sum_{\mathbf{k}s}\left[\left(q_{\mathbf{k}s}(t) + \frac{i}{\omega_\mathbf{k}}p_{\mathbf{k}s}(t)\right)\boldsymbol{\epsilon}_{\mathbf{k}s}e^{i\mathbf{k}\cdot\mathbf{r}} + \text{c.c.}\right], 
\label{Eq: 18a} \\[4pt]
\mathbf{E}(\mathbf{r},t)
&= \frac{i}{2 \sqrt{\epsilon_0 L^3}}\sum_{\mathbf{k}s}\left[\left(\omega_\mathbf{k} q_{\mathbf{k}s}(t) + i p_{\mathbf{k}s}(t)(t)\right)\boldsymbol{\epsilon}_{\mathbf{k}s}e^{i\mathbf{k}\cdot\mathbf{r}} - \text{c.c.}\right],
\label{Eq: 18b} \\[4pt]
\mathbf{B}(\mathbf{r},t)
&= \frac{i}{2 \sqrt{\epsilon_0 L^3}}
\sum_{\mathbf{k}s}\Bigg[
\left(q_{\mathbf{k}s}(t) + \frac{i}{\omega_\mathbf{k}}p_{\mathbf{k}s}(t)\right)\mathbf{k}\times\boldsymbol{\epsilon}_{\mathbf{k}s}
e^{i\mathbf{k}\cdot\mathbf{r}}  \notag\\
&\qquad\qquad\qquad
- \text{c.c.}
\Bigg].
\label{Eq: 18c}
\end{align}
\end{subequations}
Upon quantization, the canonical variables become operators $\hat{q}_{\mathbf{k}s}(t)$ and $\hat{p}_{\mathbf{k}s}(t)$ obeying
\begin{subequations}
\begin{align}
[\hat{q}_{\mathbf{k}s}(t), \hat{p}_{\mathbf{k}'s'}(t)]= & i\hbar \delta^3_{\mathbf{k}\mathbf{k}'}\delta_{ss'},\\
[\hat{q}_{\mathbf{k}s}(t), \hat{q}_{\mathbf{k}'s'}(t)]= & 0,\\
[\hat{p}_{\mathbf{k}s}(t), \hat{p}_{\mathbf{k}'s'}(t)]= & 0.
\end{align}
\end{subequations}
The Hamiltonian of the quantized radiation field is
\begin{equation}
\hat{H} = \frac{1}{2}\sum_{\mathbf{k}s} \left[\hat{p}_{\mathbf{k}s}^2(t) + \omega_\mathbf{k}^2 \hat{q}_{\mathbf{k}s}^2(t)\right].
\label{Eq: 20}
\end{equation}
The operators $\hat{q}_\mathbf{k}(t)$ and $\hat{p}_\mathbf{k}(t)$ are Hermitian. It is, however, convenient to introduce the non-Hermitian annihilation and creation operators
\begin{subequations}
\begin{align}
\hat{a}_{\mathbf{k}s} &= \frac{1}{\sqrt{2\hbar\omega_\mathbf{k}}}[\omega_\mathbf{k} \hat{q}_{\mathbf{k}s}(t) + i\hat{p}_{\mathbf{k}s}(t)], \\
\hat{a}_{\mathbf{k}s}^\dagger &= \frac{1}{\sqrt{2\hbar\omega_\mathbf{k}}}[\omega_\mathbf{k} \hat{q}_{\mathbf{k}s}(t) - i\hat{p}_{\mathbf{k}s}(t)],
\end{align}
\end{subequations}
which satisfy the standard bosonic commutation relations

\begin{subequations}
\begin{align}
[\hat{a}_{\mathbf{k}s}(t), \hat{a}_{\mathbf{k}'s'}^\dagger(t)] = &\delta_{\mathbf{k}\mathbf{k}'}^3\delta_{ss'}, \\
[\hat{a}_{\mathbf{k}s}(t), \hat{a}_{\mathbf{k}'s'}(t)] = & 0, \\
[\hat{a}_{\mathbf{k}s}^\dagger(t), \hat{a}_{\mathbf{k}'s'}^\dagger(t)] = & 0.
\end{align}
\end{subequations}
The inverse relations express the canonical variables in terms of 
$\hat{a}_{\mathbf{k}s}(t)$ and $\hat{a}_{\mathbf{k}s}^\dagger(t)$ 
\begin{subequations}
\begin{align}
\hat{q}_{\mathbf{k}s}(t) &= \sqrt{\frac{\hbar}{2\omega_k}}[\hat{a}_{\mathbf{k}s}(t) + \hat{a}_{\mathbf{k}s}^\dagger(t)], \\
\hat{p}_{\mathbf{k}s}(t) &= i\sqrt{\frac{\hbar\omega_k}{2}}[\hat{a}_{\mathbf{k}s}^\dagger(t) - \hat{a}_{\mathbf{k}s}(t)].
\end{align}
\end{subequations}
Substituting these into Eq. \eqref{Eq: 18a}-\eqref{Eq: 18c}, we get
\begin{subequations}
\begin{align}
\hat{\mathbf{A}}(\mathbf{r},t)
= & \sum_{\mathbf{k},s}
\sqrt{\frac{\hbar}{2\varepsilon_0 \omega_\mathbf{k} L^{3}}}
\left[
\hat{a}_{\mathbf{k}s}(t)\,\boldsymbol{\epsilon}_{\mathbf{k}s}\,e^{i\mathbf{k}\cdot\mathbf{r}}
+
\text{h.c}
\right]\\
\hat{\mathbf{E}}(\mathbf{r},t)
= & i\sum_{\mathbf{k},s}
\sqrt{\frac{\hbar\omega_\mathbf{k}}{2\varepsilon_0 L^{3}}}
\left[
\hat{a}_{\mathbf{k}s}(t)\,\boldsymbol{\epsilon}_{\mathbf{k}s}\,e^{i\mathbf{k}\cdot\mathbf{r}}
-
\text{h.c}
\right]\\
\hat{\mathbf{B}}(\mathbf{r},t)
= &\, i\sum_{\mathbf{k},s}
\sqrt{\frac{\hbar}{2\varepsilon_0\omega_\mathbf{k} L^{3}}}
\Big[
\hat{a}_{\mathbf{k}s}(t)\,(\mathbf{k}\times\boldsymbol{\epsilon}_{\mathbf{k}s})\,e^{i\mathbf{k}\cdot\mathbf{r}}
-\text{h.c} \Big].
\end{align}
\end{subequations}
These equations represent the quantized form of the electric and magnetic fields. Substituting the field operators into the Hamiltonian Eq. \eqref{Eq: 20} yields
\begin{equation}
\hat{H} = \sum_{\mathbf{k},s}\hbar\omega_\mathbf{k}\left(\hat{a}_{\mathbf{k}s}^\dagger \hat{a}_{\mathbf{k}s} + \frac{1}{2}\right),
\end{equation}
where each mode behaves as a quantum harmonic oscillator with zero-point energy $\frac{1}{2}\hbar\omega_\mathbf{k}$.

\subsection{Quantization of Lumped -Element LC Circuits}
\begin{figure}[ht]
    \centering
    \includegraphics[width=0.8\linewidth]{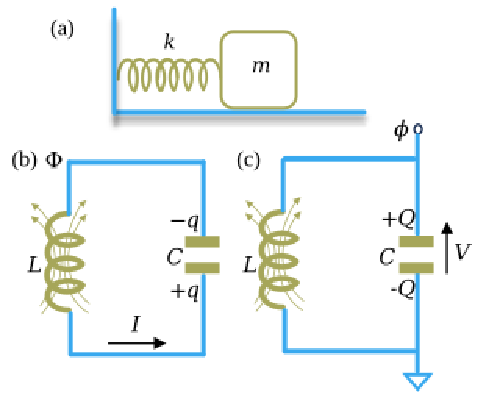}
   \caption{
    (a) The schematic diagram of a spring-mass system. (b) Schematic illustration of an LC oscillator analogous to a mechanical mass-spring system. In this representation, the charge $q$ accumulated on the capacitor due to the current $I$ flowing through the inductor plays the role of the position coordinate, while the magnetic flux $\Phi$ through the inductor acts as the conjugate momentum.
    (b) Alternative description in terms of circuit flux variables, where the generalized coordinate is the node flux $\phi = \int V\,dt$, defined as the time integral of the voltage $V$ across the capacitor, and the conjugate momentum is the charge $Q$. The charge on the capacitor arises from the electrochemical potential difference between the two plates.
    }
    \label{fig:1}
\end{figure}
Having established the quantization of the EM field, we now consider its simplest circuit realization: the lumped-element LC oscillator (where the physical size is much smaller than the wavelength of EM waves at the oscillator frequency, $\lambda = 2\pi c / \Omega_0$) shown schematically in Fig.~\ref{fig:1}(b). Each resonator represents a single quantized EM mode and is analogous to the mechanical harmonic oscillator as illustrated in Fig.~\ref{fig:1}(a). The formal equivalence between these systems provides a direct route to circuit quantization \cite{girvin2011superconducting}.
For reference, the mechanical harmonic oscillator consists of a mass $m$ attached to a spring of stiffness $k$. If $x$ denotes the displacement from equilibrium, its Lagrangian $\mathcal{L}$ is
\begin{equation}
\mathcal{L} = \frac{1}{2} m \dot{x}^2 - \frac{1}{2} k x^2 .
\end{equation}
The conjugate momentum corresponding to $x$ is defined as
\begin{equation}
p = \frac{\partial \mathcal{L}}{\partial \dot{x}} = m \dot{x},
\end{equation}
and the Hamiltonian, obtained via the Legendre transformation, takes the form
\begin{equation}
H = p\dot{x} - \mathcal{L}
= \frac{p^2}{2m} + \frac{1}{2} k x^2 .
\end{equation}
Thus, the total energy is the sum of kinetic energy $p^2/2m$ and potential energy $kx^2/2$, with the natural oscillation frequency
\begin{equation}
\Omega = \sqrt{\frac{k}{m}}.
\end{equation}
A single mode of the EM can be described in close analogy with the mechanical harmonic oscillator by considering a lumped-element LC resonator. In this case, the capacitor stores electric energy $(q^2/2C)$ due to charge separation, analogous to potential energy, while the inductor stores magnetic energy $(LI^2/2)$ associated with the current flow, analogous to kinetic energy. Here $q$ denotes the charge on the capacitor and $I$ is the current through the inductor.
 Combining these energy contributions, the Lagrangian of the LC circuit is written as
\begin{equation}
\mathcal{L} = \frac{1}{2}L I^2 - \frac{1}{2}\frac{q^2}{C}.
\end{equation}
\noindent Since the current is the time derivative of charge, $I = \dot{q}$, and $q$ plays the role of a position variable; hence, the Lagrangian turns into
\begin{equation}
\mathcal{L} = \frac{L}{2}  \dot{q}^2 - \frac{1}{2}\frac{q^2}{C}.
\end{equation}
 From the Euler–Lagrange equation, one can obtains the equation of motion
 \begin{equation}
     \frac{d}{dt}\Big(\frac{\partial \mathcal{L}}{\partial \dot{q}}\Big) - \frac{\partial \mathcal{L}}{\partial q} = 0\implies \ddot{q} + \Omega^2 q = 0,
 \end{equation}
 with the natural resonance frequency, $\Omega = 1/\sqrt{LC}$. The conjugate momentum variable to $q$ is the flux $\Phi$ through the inductor, defined as
 \begin{equation}
     \Phi = \frac{\partial \mathcal{L}}{\partial \dot{q}} = L\dot{q} = LI.
 \end{equation}
 Thus, the generalized coordinates of the LC oscillator are $(q, \Phi)$, directly analogous to the position–momentum pair $(x,p)$ in mechanics.
 The Hamiltonian of the system can be written as
 \begin{equation}
     H = \Phi \dot{q} - \mathcal{L} = \frac{\Phi^2}{2L} + \frac{q^2}{2C}.
 \end{equation}
This Hamiltonian is formally identical to that of the mechanical harmonic oscillator under the correspondence
\vspace{0.4cm}
\begin{ruledtabular}
\begin{tabular}{cc}
\hline
\textbf{Mechanical Oscillator} & \textbf{LC Oscillator} \\
\hline
$x$ & $q$ \\
$p$ & $\Phi$ \\
$m$ & $L$ \\
$k$ & $1/C$ \\
$\Omega = \sqrt{k/m}$ & $\Omega = 1/\sqrt{LC}$ \\
\hline
\end{tabular}
\end{ruledtabular}
\vspace{0.2cm}
Thus, the LC resonator represents the electrical realization of a harmonic oscillator. Since both systems possess an identical quadratic Hamiltonian structure, quantization leads directly to the same operator algebra. This analogy forms the theoretical foundation of superconducting cQED, where engineered LC modes are coupled to artificial atoms \cite{PhysRevA.69.062320}.
From this Hamiltonian, the equations of motion follow directly from Hamilton’s canonical relations. 
 \begin{subequations}
\begin{align}
    \dot{q} &= \frac{\partial H}{\partial \Phi} = \frac{\Phi}{L} = I, \label{eq:qdot}\\
    \dot{\Phi} &= -\frac{\partial H}{\partial q} = -\frac{q}{C} = V. \label{eq:phidot}
\end{align}
\end{subequations}
It gives the current flowing through the inductor and the voltage across the capacitor at the node connecting the two elements.
The coordinate and its conjugate, momentum, can now be written with quantum operators obeying the commutation relation
\begin{equation}
    [\hat{\Phi}, \hat{q}] = -i\hbar,
\end{equation}
The Hamiltonian then takes the familiar harmonic oscillator form,
\begin{equation}
    H = \hbar \Omega (\hat{a}^\dagger \hat{a} + \frac{1}{2}),
\end{equation}
where the annihilation and creation operators are defined as
\begin{subequations}
\begin{align}
    \hat{a} = & i\frac{1}{\sqrt{2L\hbar \Omega}}\hat{\Phi} + \frac{1}{\sqrt{2C\hbar \Omega}}\hat{q}\\
    \hat{a}^\dagger = & -i\frac{1}{\sqrt{2L\hbar \Omega}}\hat{\Phi} + \frac{1}{\sqrt{2C\hbar \Omega}}\hat{q}
\end{align}
\end{subequations}
which satisfy the commutation relation $[\hat{a}, \hat{a}^\dagger] = 1$.
We have successfully derived the Hamiltonian for LC resonator, which resembles the quantum harmonic oscillator with $q$ as our coordinate of choice and $\Phi$ as its conjugate momentum. However, in the context of the Josephson junction, our LC resonator will have a nonlinear inductance, so it will be more convenient to choose the node flux as the coordinate basis. Let us define the node flux at the point as shown in Fig. \ref{fig:1}(c) by
\begin{equation}
    \phi(t) = \int^t_{-\infty} V(\tau)\, d\tau,
    \label{eq:phi_def}
\end{equation}
so that \( V(t) = \dot{\phi} \)\cite{Devoret1997}. Then the potential energy stored on the capacitor is
\begin{equation}
    U = \frac{1}{2} C \dot{\phi}^2,
    \label{eq:cap_energy}
\end{equation}
and now looks like the kinetic energy with this choice of coordinate. Similarly, using Faraday’s law and the sign convention for the direction of the current defined in Fig. \ref{fig:1}(c) we have
\begin{equation}
    V = L \dot{I} = \dot{\phi},
    \label{eq:ind_voltage}
\end{equation}
 and thus see that the node flux variable \(\phi\) really is the physical magnetic flux \(\Phi\) winding through the inductor (ignoring any possible external flux applied through the loop of the circuit or the inductor). Hence, the kinetic energy stored in the inductor is
\begin{equation}
    T = \frac{1}{2L} \phi^2,
    \label{eq:ind_energy}
\end{equation}
which now looks like the potential energy. With this choice of coordinate, the Lagrangian becomes
\begin{equation}
    \mathcal{L} = \frac{1}{2} C \dot{\phi}^2 - \frac{1}{2L} \phi^2,
    \label{eq:lagrangian}
\end{equation}
and the momentum conjugate to the flux is
\begin{equation}
    Q = \frac{\delta \mathcal{L}}{\delta \dot{\phi}} = C \dot{\phi},
    \label{eq:charge}
\end{equation}
which is now the charge as defined with the sign convention in Fig. \ref{fig:1}(c). Notice the crucial minus-sign relative to the previous result. This is necessary to maintain the sign of the commutation relation when we interchange the momentum and coordinate.  To reiterate: when the charge is the coordinate and the flux is the conjugate momentum, the commutation relation is
\begin{equation}
    [\hat{q}, \hat{\Phi}] = + i \hbar.
    \label{eq:commutation}
\end{equation}
 whereas when the flux is the coordinate and the charge is the conjugate momentum, the commutation relation is:
 \begin{equation}
    [\hat{\phi}, \hat{Q}] = + i \hbar.
    \label{eq:commutation}
\end{equation}
Since we have chosen a convention in which $\hat{\Phi} = \hat{\phi}$ and $\hat{Q} = -\hat{q}$
Just to be completely explicit, we now repeat the derivation of the Hamiltonian and
its quantization for this new choice which we will be using throughout the remainder
of these notes. Thus the Hamiltonian can be written
\begin{equation}
    H = Q \dot{\phi} - \mathcal{L} = \frac{Q^2}{2C} + \frac{\phi^2}{2L}.
    \label{eq:Hamiltonian_flux_charge}
\end{equation}
Hamilton’s equations of motion are then
\begin{align}
    \dot{\phi} &= +\frac{\partial H}{\partial Q} = +\frac{Q}{C}, \label{eq:Hamilton_eq_phi}\\
    \dot{Q} &= -\frac{\partial H}{\partial \phi} = -\frac{\phi}{L}. \label{eq:Hamilton_eq_Q}
\end{align}
Again in the usual way, the coordinate and its conjugate momentum can be promoted
to quantum operators obeying the canonical commutation relation
\begin{equation}
    [\hat{Q}, \hat{\phi}] = -i\hbar.
    \label{eq:commutation_flux_charge}
\end{equation}
and we can write the Hamiltonian
\begin{equation}
    H = \frac{\hbar \Omega}{2} \left\{ \hat{a}^\dagger \hat{a} + \hat{a} \hat{a}^\dagger \right\}
    = \hbar \Omega \left\{ \hat{a}^\dagger \hat{a} + \frac{1}{2} \right\},
    \label{eq:Hamiltonian_ladder}
\end{equation}
in terms of raising and lowering operators
\begin{align}
    \hat{a} &= +i \frac{1}{\sqrt{2C\hbar\Omega}} \hat{Q}
            + \frac{1}{\sqrt{2L\hbar\Omega}} \hat{\phi}, \label{eq:a_operator}\\
    \hat{a}^\dagger &= -i \frac{1}{\sqrt{2C\hbar\Omega}} \hat{Q}
            + \frac{1}{\sqrt{2L\hbar\Omega}} \hat{\phi}, \label{eq:adag_operator}
\end{align}
which obey the usual relation
\begin{equation}
    [\hat{a}, \hat{a}^\dagger] = 1.
    \label{eq:comm_a_adag}
\end{equation}
The charge and flux operators can be expressed in terms of the raising and lowering
operators as
\begin{align}
    \hat{Q} &= -i Q_{\text{ZPF}} \left( \hat{a} - \hat{a}^\dagger \right), \label{eq:Q_operator}\\
    \hat{\phi} &= \Phi_{\text{ZPF}} \left( \hat{a} + \hat{a}^\dagger \right), \label{eq:phi_operator}
\end{align}
where
\begin{align}
    Q_{\text{ZPF}} &= \sqrt{\frac{C \hbar \Omega}{2}} = \sqrt{\frac{\hbar}{2Z}}, \label{eq:QZPF}\\
    \Phi_{\text{ZPF}} &= \sqrt{\frac{L \hbar \Omega}{2}} = \sqrt{\frac{\hbar Z}{2}}, \label{eq:PhiZPF}
\end{align}
where \( Z \) is the characteristic impedance of the oscillator.
\begin{equation}
    Z = \sqrt{\frac{L}{C}}.
    \label{eq:Z}
\end{equation}
Notice that the notation has been chosen such that the quantum ground state
uncertainties in charge and flux are given by
\begin{align}
    \langle 0 | \hat{Q}^2 | 0 \rangle &= Q_{\text{ZPF}}^2, \label{eq:Q_uncertainty}\\
    \langle 0 | \hat{\phi}^2 | 0 \rangle &= \Phi_{\text{ZPF}}^2. \label{eq:phi_uncertainty}
\end{align}
Using the superconducting resistance quantum
\begin{equation}
    R_Q \equiv \frac{h}{(2e)^2} \approx 6{,}453.20 \hspace{0.1cm}\text{Ohms}, 
    \label{eq:RQ}
\end{equation}
we can define a dimensionless characteristic impedance
\begin{equation}
    z \equiv {Z}/{R_Q},
    \label{eq:z_def}
\end{equation}
to obtain
\begin{subequations}
\begin{align}
    Q_{\text{ZPF}} &= (2e)\sqrt{\frac{1}{4\pi z}}, \label{eq:Qzpf2}\\
    \Phi_{\text{ZPF}} &= \Phi_0 \sqrt{\frac{z}{4\pi}}, \label{eq:Phizpf2}
\end{align}
\end{subequations}
where
\begin{equation}
    \Phi_0 \equiv \frac{h}{2e}
    \label{eq:Phi0}
\end{equation}
is the superconducting flux quantum. Notice that the usual uncertainty product is obeyed:
\begin{equation}
    Q_{\text{ZPF}} \Phi_{\text{ZPF}} = \frac{\hbar}{2}.
    \label{eq:uncertainty}
\end{equation}
Since a lumped-element LC resonator represents a single quantized mode, many experimentally relevant systems support spatially distributed excitations. A microwave transmission line constitutes the simplest extension of the LC oscillator to a system with infinitely many coupled degrees of freedom. Physically, it can be viewed as a continuum limit of discrete inductors and capacitors arranged along a propagation direction, thereby forming a one-dimensional EM environment \cite{PhysRevA.69.062320}.

In contrast to the lumped resonator, where the dynamical variable depends only on time, the transmission line is described by a flux field that varies in both space and time. The quantization procedure therefore parallels that of a continuous field, closely resembling the normal-mode decomposition of the EM discussed earlier.

\subsection{Quantization of Microwave Transmission Line}
\begin{figure}[h]
    \centering
 \includegraphics[width=\linewidth]{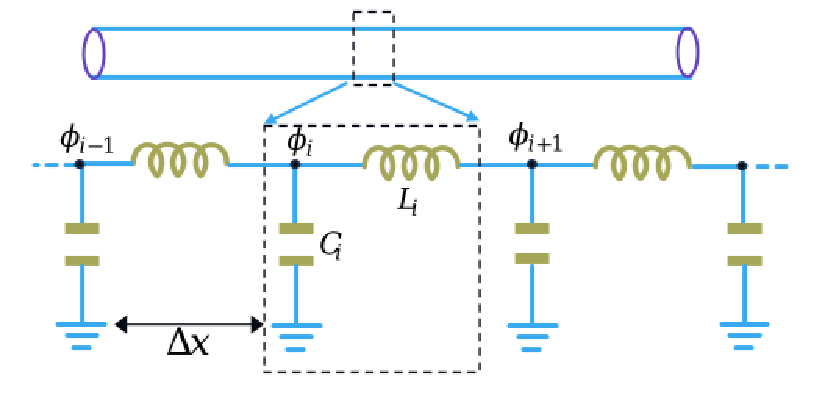}
    \caption{Schematic of transmission line quantization, each infinitesimal section is modeled as an LC oscillator with distributed inductance and capacitance.}
    \label{fig:3}
\end{figure}

We model the transmission line as a one-dimensional array of infinitesimal circuit elements \cite{RevModPhys.93.025005,pozar2011microwave}, as illustrated in Fig.~\ref{fig:3}. Each segment of length $\Delta x$ possesses an inductance and capacitance determined by the inductance and capacitance per unit length, $l(x)$ and $c(x)$:
\begin{equation}
    C_i = c(x_i)\, \Delta x, 
    \qquad 
    L_i = l(x_i)\, \Delta x.
\end{equation}
The Lagrangian of this system is
\begin{equation}
    \mathcal{L} = \frac{1}{2} \sum_{i=1}^N C_i \dot{\Phi}_i^2 - \sum_{i=1}^{N-1} \frac{(\Phi_{i+1} - \Phi_i)^2}{2L_i},
\end{equation}
where $\Phi_i$ is the node flux. The conjugate momentum is:
\begin{equation}
    Q_i = \frac{\partial \mathcal{L}}{\partial \dot{\Phi}_i} = C_i \dot{\Phi}_i,
\end{equation}
giving the Hamiltonian:
\begin{equation}
    \mathcal{H} = \sum_{i=1}^{N} \frac{Q_i^2}{2C_i} + \sum_{i=1}^{N-1} \frac{(\Phi_{i+1} - \Phi_i)^2}{2L_i}.
\end{equation}
To connect this discrete model to the continuous transmission line, take the limit $\Delta x \to 0$. In this limit, identify
\begin{equation}
    \Phi_{i+1} - \Phi_i \approx \Delta x\, \partial_x \Phi(x_i), 
    \qquad 
    Q_i = \rho(x_i)\, \Delta x,
\end{equation}
where $\rho(x)$ is the charge density. The Hamiltonian becomes 
\begin{equation}
    \mathcal{H} = \int_0^L dx \left[ \frac{\rho(x)^2}{2c(x)} + \frac{1}{2l(x)} (\partial_x \Phi(x))^2 \right],
\end{equation}
with canonical commutation relation $[\Phi(x), \rho(x')] = i \hbar \delta(x-x')$. For simplicity, take $c(x) = c$, $l(x) = l$. The continuous Lagrangian is 
\begin{equation}
    \mathcal{L}_g = \int_0^L dx \frac{1}{2} \Big[ c (\partial_t \Phi)^2 - \frac{1}{l} (\partial_x \Phi)^2 \Big].
\end{equation}
The Euler–Lagrange equation gives the wave equation
\begin{equation}
    \partial_t^2 \Phi - v_p^2 \partial_x^2 \Phi = 0, \quad v_p = \frac{1}{\sqrt{lc}}.
\end{equation}
The conjugate momentum is
\begin{equation}
    Q(x,t) = \frac{\partial \mathcal{L}_g}{\partial \dot{\Phi}} = c \dot{\Phi}.
\end{equation}
Solve the wave equation using $\Phi(x,t) = e^{-i\omega t} \phi(x)$, gives
\begin{equation}
    v_p^2 \partial_x^2 \phi + \omega^2 \phi = 0.
\end{equation}
with $\omega = v_p k$. For open boundary conditions, the eigenfunctions are
\begin{equation}
    \phi_n(x) = \sqrt{\frac{2}{L}} \cos(k_n x), \quad k_n = \frac{n\pi}{L}, \quad n=0,1,2,\dots
\end{equation}
with the orthogonality condition
\begin{equation}
\int_0^L \phi_n \phi_m dx = \delta_{nm}, \quad 
\int_0^L (\partial_x \phi_n)(\partial_x \phi_m) dx = k_n^2 \delta_{nm}.
\end{equation}
Expand the flux operator as
\begin{equation}
    \Phi(x,t) = \sum_{n=0}^\infty \zeta_n(t) \phi_n(x),
\end{equation}
giving the Lagrangian in normal modes
\begin{equation}
    \mathcal{L}_g = \frac{1}{2} \sum_n \Big[ c (\dot{\zeta}_n)^2 - c \omega_n^2 \zeta_n^2 \Big], \quad \omega_n = v_p k_n.
\end{equation}
The conjugate momentum is $Q_n = c \dot{\zeta}_n$, 
and the Hamiltonian becomes:
\begin{equation}
    H = \frac{1}{2} \sum_n \left[ \frac{Q_n^2}{c} + c \omega_n^2 \zeta_n^2 \right].
\end{equation}
The $n=0$ mode corresponds to a uniform charge distribution along the line and behaves as a free particle rather than a harmonic oscillator. Since this mode contributes only a constant total charge, it is typically neglected. The remaining modes are quantized by introducing bosonic creation and annihilation operators:
\begin{subequations}
\begin{align}
    \hat{\zeta}_n = \sqrt{\frac{\hbar}{2c \omega_n }}(a_n + a_n^\dagger),\\
    \hat{Q}_n = -i \sqrt{\frac{\hbar c \omega_n }{2}}(a_n - a_n^\dagger).
\end{align}
\end{subequations}
The flux and charge operators in real space are then
\begin{equation}
    \hat{\Phi}(x) = \sum_n \hat{\zeta}_n \phi_n(x), \quad
    \hat{Q}(x) = \sum_n \hat{Q}_n \phi_n(x),
\end{equation}
with $[\hat{\Phi}(x), \hat{Q}(x')] = i \hbar \delta(x-x')$. 
Finally, the quantized Hamiltonian is:
\begin{equation}
    \hat{H} = \int_0^L dx \left[ \frac{\hat{Q}^2}{2c} + \frac{1}{2l} (\partial_x \hat{\Phi})^2 \right] 
    = \sum_n \hbar \omega_n \left(a_n^\dagger a_n + \frac{1}{2} \right),
\end{equation}
describing an infinite set of harmonic modes, the standing EM waves along the line.
\subsection{Dissipation in Quantum Systems}
% \subsection{Open Quantum Systems and the Lindblad Master Equation}
Up to this stage, we have developed a fully quantized description of EM degrees of freedom, ranging from free-space field modes to lumped-element LC resonators and distributed transmission lines. In each case, the classical EM was promoted to an operator, leading to discrete harmonic excitations that represent quantized light.

\begin{figure}[ht]
    \centering
    \includegraphics[width=0.4\linewidth]{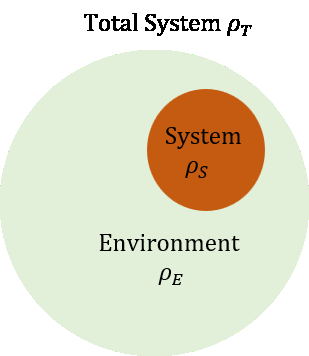}
    \caption{
    Schematic representation of an open quantum system. The total system, described by the density matrix $\rho_T$, consists of a subsystem of interest (system, $\rho_S$) interacting with its surrounding environment (environment, $\rho_E$). Coupling between the system environment leads to the exchange of energy and information, giving rise to open-system dynamics.
    }

    \label{fig:2}
\end{figure}

In realistic quantum platforms, quantized EM modes interact with material systems such as atoms, artificial atoms, or superconducting qubits, whose energy spectra are inherently discrete. It is important to emphasize that constructing a perfectly isolated quantum system is practically impossible. The field modes and matter (system) degrees of freedom inevitably couple to their surrounding environment, as shown in Fig. \ref{fig:2}. This unavoidable interaction leads to energy relaxation, decoherence, and dissipation. Consequently, unlike idealized closed systems governed solely by unitary dynamics, real quantum systems must be described within the framework of open quantum system theory \cite{breuer2002theory, scully1997quantum}. Therefore, before introducing coherent light–matter coupling, it is essential to establish a formalism capable of consistently incorporating dissipation and environmental effects.
In quantum mechanics, the time evolution of a closed system governed by a Hamiltonian $H$ is unitary and determined by the Schrödinger equation,
\begin{equation}
i\hbar \frac{d}{dt} |\psi(t)\rangle = H |\psi(t)\rangle.
\end{equation}
In the density operator formalism, this evolution is equivalently written by the von Neumann equation \cite{von2018mathematical}.
\begin{equation}
\frac{d\rho}{dt} = -\frac{i}{\hbar}[H,\rho].
\end{equation}
This unitary dynamics preserves trace, positivity, and purity, ensuring reversible evolution within an isolated system.
To incorporate dissipation, we consider a composite description in which the system of interest is coupled to an environment. The total Hilbert space is expressed as
\begin{equation}
\mathcal{H}_\text{T} = \mathcal{H}_\text{S} \otimes \mathcal{H}_\text{E},
\end{equation}
where $\mathcal{H}_\text{S}$ denotes the subsystem and $\mathcal{H}_\text{E}$ the environmental degrees of freedom, shown in Fig.~\ref{fig:2}. The total density operator $\rho_\text{T}(t)$ evolves unitarily under the total Hamiltonian
\begin{equation}
H_\text{T} = H_\text{S} + H_\text{E} + H_\text{I},
\end{equation}
 where, $H_\text{S}$ is the system Hamiltonian. $H_\text{E}$ is the environment Hamiltonian, and $H_\text{I}$ is the interaction Hamiltonian. 
Although the combined system environment state evolves unitarily, experimental access is typically restricted to the subsystem alone. The effective state of the subsystem is therefore obtained by tracing out the environmental degrees of freedom.
\begin{equation}
\rho_\text{S}(t) = \mathrm{Tr}_E[\rho_\text{T}(t)].
\end{equation}
The resulting reduced dynamics need not be unitary, even though the total evolution is unitary.
Under the assumptions of weak coupling, short bath memory, and Markovian behavior,  the evolution of $\rho_\text{S}(t)$ is governed by the Lindblad master equation \cite{breuer2002theory},
\begin{equation}
\dot{\rho}_\text{S}(t)
=
-\frac{i}{\hbar}[H_\text{S},\rho_\text{S}(t)]
+
\mathcal{L}[\rho_\text{S}(t)],
\end{equation}
where $\hat{\mathcal L}$ is a Lindblad operator. It describes the nature of the energy loss or decoherence process causing the dynamics to be irreversible.
The dissipative superoperator $\mathcal{L}$ can be written in the standard form
\begin{equation}
\mathcal{L}[\rho_\text{S}]
=
\sum_{i}
\frac{\gamma_i}{2}
\left(
2L_i \rho_\text{S} L_i^\dagger
-
L_i^\dagger L_i \rho_\text{S}
-
\rho_\text{S} L_i^\dagger L_i
\right),
\end{equation}
where $\gamma_i$ denotes the decay rate associated with a given dissipative channel and $L_i$ are the corresponding jump operators. 
For radiative decay between atomic energy levels, the jump operators reduce to lowering operators $L_i = \sigma_i^{-} = |j\rangle\langle i|$, describing transitions from an excited state $|i\rangle$ to a lower state $|j\rangle$. The Lindblad formalism thus provides a consistent and physically well-defined description of energy relaxation and decoherence in quantum optical and solid-state platforms.

Having established the open-system framework required to describe dissipation, we now turn to the simplest quantum emitter, the two-level system, which serves as the fundamental building block for modeling quantized matter and light–matter interactions.
\subsection{Quantum Coherence in Two-Level Quantum Systems}
\begin{figure}
    \centering
    \includegraphics[width=\linewidth]{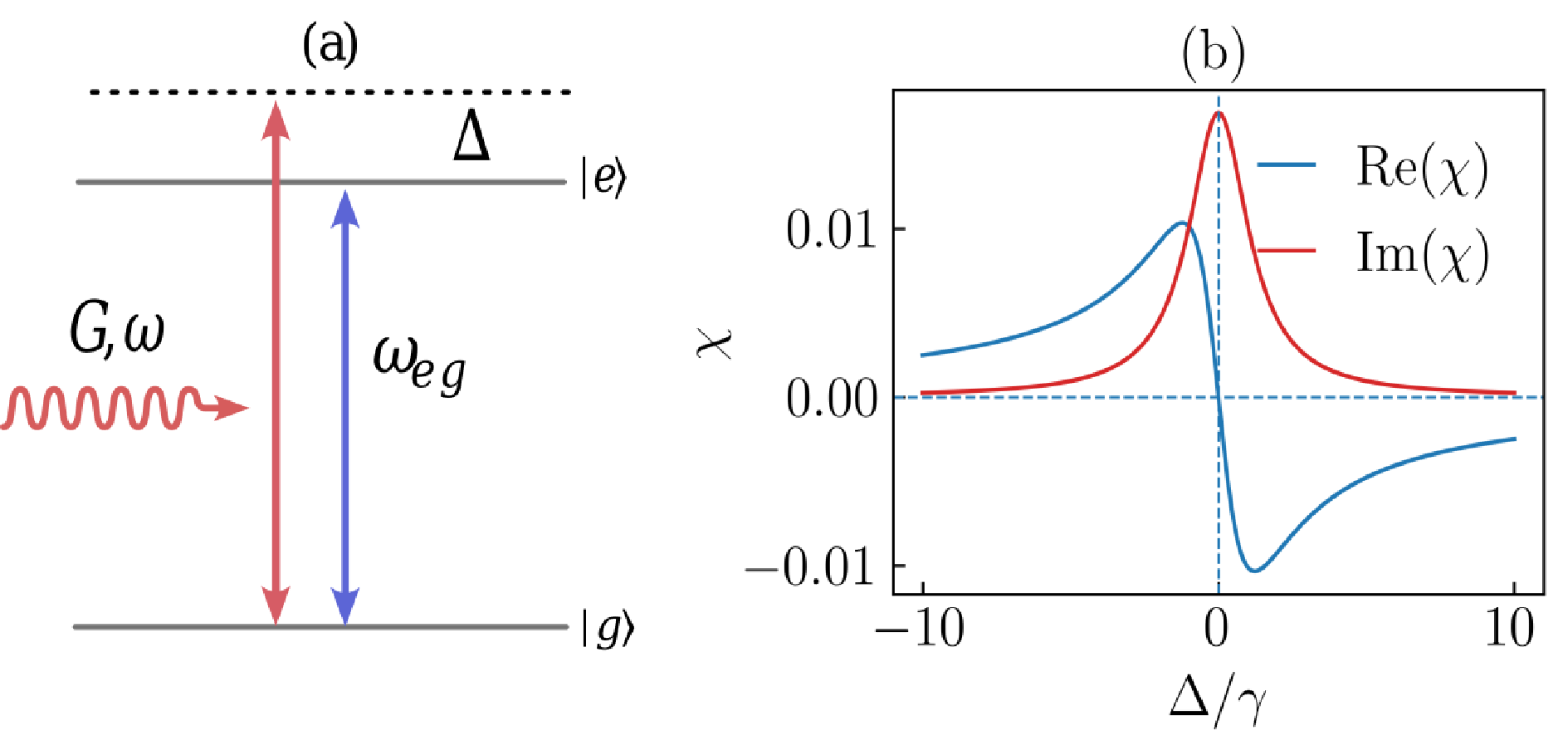}
    \caption{
(a) Energy level diagram of a two-level atom with ground state $|g\rangle$ and excited state $|e\rangle$, separated by the transition frequency $\omega_{eg}$. The atom interacts with a monochromatic EM of frequency $\omega$ with Rabi frequency $G$. The detuning is defined as $\Delta = \omega - \omega_{eg}$. 
(b) Real part $\chi'(\omega)$ and imaginary part $\chi''(\omega)$ of the complex electric susceptibility $\chi$ as a function of detuning $\Delta$. The imaginary part represents absorption and exhibits a Lorentzian profile centered at $\Delta = 0$, while the real part represents dispersion and shows a dispersive line shape. The parameters used in the plots are: 
$\lambda = 780\,\text{nm}$, 
$\gamma = 3.81 \times 10^{7}\,\text{s}^{-1}$, 
$N = 10^{18}\,\text{m}^{-3}$, 
and $G = 0.5\,\gamma$..
}
    \label{fig:2level}
\end{figure}

A two-level system is the simplest quantum mechanical model of matter. While real atoms possess many energy levels, under resonant excitation at a particular frequency, the dynamics can often be confined to just two dominant states: ground state $|g \rangle$ and excited state $|e\rangle$, while all other levels are highly detuned. In such cases, the atom can be effectively treated as a two-level system, as shown in Fig. \ref{fig:2level}(a). The respective eigenvalues of the states $|g\rangle$ and $|e \rangle$ are $\hbar \omega_g$ and  $\hbar \omega_e$  for the unperturbed
Hamiltonian $H_0$. By using the completeness relation $|g\rangle \langle g| + |e\rangle \langle e| = 1$, we write the
Hamiltonian $H_0$ as 
\begin{equation}
    H_0 = \hbar \omega_{g} |g\rangle \langle g| + \hbar \omega_{e} |e\rangle \langle e|,
\end{equation}
We set the ground-state energy as the reference level, $\omega_g =0$. Introducing the transition frequency $\omega_{eg} = \omega_e - \omega_g$, the Hamiltonian $H_0$ simplifies to
\begin{equation}
H_0
= \hbar \omega_{eg} |e\rangle \langle e|.
\end{equation}
In the presence of a quasi-monochromatic radiation, the interaction Hamiltonian of the atom under the dipole moment approximation (the wavelength of interacting EM have much larger than the typical size of an atom) is given by
\begin{equation}
     H_\text{I} =  -\mathbf{d}.\mathbf{E(r,t)},
\end{equation}
where $\mathbf{d}$ is the dipole matrix and $\mathbf{E(r,t)}$ is the incident plane monochromatic EM field, has the form
\begin{subequations}
\begin{align}
\mathbf{d} &= \mathbf{d}_{eg} |e \rangle \langle g| 
+ \mathbf{d}_{ge} |g \rangle \langle e|, \\
\mathbf{E}(\mathbf{r},t) &= \hat{\mathbf{e}}\,\mathcal{E}_0(\mathbf{r}) 
e^{i(\mathbf{k}\cdot \mathbf{r} - \omega t)} 
+ \mathrm{c.c.}
\end{align}
\end{subequations}
The off-diagonal elements $\mathbf{d}_{eg}$ and $\mathbf{d}_{ge}$ represent the induced dipole moments of the atom. The diagonal elements, $\mathbf{d}_{ee}$ and $\mathbf{d}_{gg}$, vanish because the dipole operator $\mathbf{d}$ is an odd-parity operator. Therefore, the elements of the dipole operator $\mathbf{d}$ will be non-zero if and only if the states $|e \rangle$ and $|g \rangle$ have different parity. Here, $\hat{e}$, $\mathcal{E}_0(r)$, $\mathbf{k}$, and $\omega$ are the polarization unit vector, slowly varying amplitude, wave vector, and frequency of the field, respectively. 
Substituting the field and dipole moment expression in the interaction Hamiltonian
\begin{equation}
\begin{aligned}
    H_\text{I} 
    = & -[\mathbf{d}_{eg}.\hat{e}\mathcal{E}_0(r) e^{i(\mathbf{k}\cdot \mathbf{r} - \omega t)}  + \mathbf{d}_{eg}.\hat{e}\mathcal{E}_0^*(r) e^{-i(\mathbf{k}\cdot \mathbf{r} - \omega t)} ]|e\rangle \langle g|\\
    & -[\mathbf{d}_{ge}.\hat{e}\mathcal{E}_0(r) e^{i(\mathbf{k}\cdot \mathbf{r} - \omega t)}  + \mathbf{d}_{ge}.\hat{e}\mathcal{E}_0^*(r) e^{-i(\mathbf{k}\cdot \mathbf{r} - \omega t)} ]|g\rangle \langle e|,
\end{aligned}
\end{equation}
The Hamiltonian can be reshaped into
\begin{equation}
\begin{aligned}
    H_\text{I} 
    = & -\hbar[(G e^{-i \omega t} + g e^{i \omega t}) |e\rangle \langle g| + \text{h.c}],
\end{aligned}
\end{equation}
where $G$ and $g$ are known as Rabi frequencies and they are defined as
\begin{equation}
   G =\frac{\mathbf{d}_{eg}.\hat{e}}{\hbar}\mathcal{E}_0(r) e^{i \mathbf{k}\cdot \mathbf{r}}, \quad g = \frac{\mathbf{d}_{eg}.\hat{e}}{\hbar}\mathcal{E}_0^*(r) e^{-i \mathbf{k}\cdot \mathbf{r}}.
\end{equation}
To make the Hamiltonian time-independent, we perform the following unitary transformation to get $H_\text{eff}$.
\begin{equation}
    H_\text{eff} = U^\dagger H_\text{T} U - i \hbar U^\dagger \frac{\partial U}{\partial t},
\end{equation}
where, $U = e^{-i \omega |e\rangle \langle e| t}$, and $H_\text{T} = H_0 + H_\text{I}$ is the total Hamiltonian.
\begin{equation}
    H_\text{eff}/\hbar = - \Delta |e\rangle \langle e| - G |e\rangle \langle g| - G^* |g\rangle \langle e|. 
\end{equation}
Here $\Delta = \omega - \omega_{eg}$ is the detuning of the incident field. To write the Hamiltonian above, we have dropped the highly oscillatory term rotating at the frequency $2\omega$, related to $g$. 
The value of  $g$ becomes important only when $g \approx \omega$. Therefore, the term $g$ can be neglected
at the optical frequency domain, where $g << 2\omega$.  This approximation is known as the rotating wave approximation (RWA) \cite{fradkin1991p}.
The state vector of $\psi$ of a two-level atom can be written as a linear combination of the excited state $|e \rangle$ and ground state $|g \rangle$
\begin{equation}
    |\psi \rangle = c_1 |g \rangle + c_2 |e \rangle,
\end{equation}
where $c_i  \{i = g,e \}$ is the probability amplitude of being in a state $|i\rangle$. The density matrix operator is defined as the projector $\rho = |\psi \rangle \langle \psi|$, and given by in matrix form as
\begin{equation}
    \rho =  \begin{pmatrix}
        \rho_{gg} & \rho_{ge} \\
       \rho_{eg} & \rho_{ee}
    \end{pmatrix} = \begin{pmatrix}
        |c_1|^2 & c_1 c_2^* \\
        c_1^* c_2 & |c_2|^2
    \end{pmatrix}.
\end{equation}
The diagonal density matrix elements $\rho_{ee}$ and $\rho_{gg}$ correspond to the populations of the excited and ground states, respectively. The off-diagonal density matrix element $\rho_{ge}$ and $\rho_{eg}$ gives the induced coherence between $|e\rangle$ and $|g\rangle$.
To study the evolution of population and coherence, let's use the Lindblad master equation
\begin{equation}
    \frac{\partial \rho(t)}{\partial t} 
    = -\frac{i}{\hbar}[H_\text{eff}, \rho] 
    + \frac{\gamma}{2} \left( 2\sigma^- \rho \sigma^+ 
    -  \{ \sigma^+ \sigma^-, \rho \} \right),
\end{equation}
where, $\{ \sigma^+ \sigma^-, \rho \} = \sigma^+ \sigma^- \rho + \rho \sigma^+ \sigma^-$ is the anticommutation relation, $\gamma$ is the decay rate of an atom in a vacuum, $\sigma^- = |g\rangle \langle e|$ and $\sigma^+ = |e\rangle \langle g|$ are lowering and raising operators.
It gives the sets of equations as
\begin{subequations}
\begin{align}
    \dot{\rho}_{ee} = &  \quad -2 \gamma \rho_{ee} + iG\rho_{ge} -iG^* \rho_{eg},\\
    \dot{\rho}_{gg} = & \quad 2 \gamma \rho_{ee} - iG\rho_{ge}  + iG^* \rho_{eg},\\
    \dot{\rho}_{eg} = & \quad  -(\gamma - i \Delta)\rho_{eg} + iG(\rho_{gg} - \rho_{ee}).
\end{align}
\end{subequations}
where $2\gamma (1/T_1)$ describes the decay rate of the atomic excited state $|e \rangle$ and $\gamma (1/T_2)$  is the
decay rate of the atomic coherence. The density matrix elements $\rho_{ee}$, $\rho_{gg}$ are related to the population conservation equation as follows: $\rho_{ee} + \rho_{gg} =1$. The off-diagonal elements $\rho_{eg}$ and $\rho_{ge}$  are complex conjugates of each other: $\rho_{ge} = \rho_{eg}^*$.
In the absence of these decays, the solution
for the atomic population and coherence (assuming initially all the atoms are in the ground state) are
\begin{equation}
\rho_{gg}
=\cos^2\!\left(\frac{\Omega t}{2}\right)
+ \frac{\Delta^2}{\Omega^2}
\sin^2\!\left(\frac{\Omega t}{2}\right),
\label{eq:rho_gg}
\end{equation}
\begin{equation}
\rho_{eg}
= \frac{2G}{\Omega^2}
\sin\!\left(\frac{\Omega t}{2}\right)
\left\{\Delta \sin\!\left(\frac{\Omega t}{2}\right)
+ i \Omega \cos\!\left(\frac{\Omega t}{2}\right)
\right\},
\label{eq:rho_eg}
\end{equation}
where $\Omega = \sqrt{\Delta^2 + 4|G|^2}$ is called the generalized Rabi frequency. For the zero detuning case, with $\omega = \omega_{eg}$, Eq. \ref{eq:rho_gg} reduces to
\begin{equation}
\rho_{gg}
=
\cos^2\!\left(\frac{\Omega t}{2}\right).
\end{equation}
The atom oscillates symmetrically between its ground and excited states with angular frequency $\Omega$. An increase in the detuning $\Delta$ leads to Rabi oscillations with reduced amplitude \cite{allen1975optical}. In the presence of relaxation, the solutions of the density matrix equations are no longer purely oscillatory. The system now settles down into a steady state after a sufficiently long time ($t >> 1/\gamma)$. In the steady-state limit, setting the time derivatives of the density matrix elements to zero yields
\begin{subequations}
\begin{align}
    \rho_{ee} = & \frac{|G|^2}{(\gamma^2 +  \Delta^2 + 2|G|^2)},\\
     \rho_{eg} = & \frac{iG(\gamma + i \Delta)}{(\gamma^2 +  \Delta^2 + 2|G|^2)}.
\end{align}
\end{subequations}
 The induced polarization in the medium at frequency $\omega$ is given by
 \begin{equation}
     \mathbf{P} =  \epsilon_0 \chi \mathbf{E} = \mathcal{N} \text{Tr}(\mathbf{d\rho}) =   \mathcal{N}({\mathbf{d}}_{ge} \rho_{eg} + {\mathbf{d}}_{eg} \rho_{ge}),
\end{equation}
 where $\mathcal{N}$ is the atomic number density. Putting the field equation, using the steady state value of $\rho_{eg}$, one can get the expression for linear susceptibility as
 \begin{equation}
     \chi = \frac{\mathcal{N}|\mathbf{d}_{eg}|^2}{\hbar \epsilon_0} \frac{(-\Delta + i \gamma)}{(\gamma^2 + \Delta^2 + 2|G|^2)}.
\end{equation}
The complex susceptibility $\chi = \chi' + i\chi''$ characterizes the linear optical response of the atomic medium to the applied field. The real part $\chi'$ describes the dispersive response of the medium and governs the refractive index and phase velocity of light propagation, while the imaginary part $\chi''$ represents the absorptive component associated with energy dissipation due to spontaneous emission. Near resonance, the imaginary part exhibits a Lorentzian absorption profile.  The full width at half maximum (FWHM) of the Lorentzian profile is $\gamma_c = \sqrt{\gamma^2 + 2|G|^2}$. The width depends on the intensity of the applied field, as shown in Fig. \ref{fig:2level}(b). Whereas the real part of $\chi$ displays a dispersive line shape, the dispersion is anomalous, and its slope determines the group velocity of light. The interplay between detuning $\Delta$, decay rate $\gamma$, and driving strength $G$ controls both absorption and dispersion, forming the basis for coherent optical phenomena such as saturation absorption, nonlinear phase modulation \cite{boyd2008nonlinear}, slow light, and transparency windows arising from CPO in the presence of a bichromatic driving field \cite{AgarwalDey2009SlowLight}.
\section{Superconducting Qubits}\label{Sec:III}
Superconducting qubits are one of the leading platforms for realizing quantum computation and quantum information processing, and on-chip quantum devices \cite{RN20919, devoret2013superconducting, RevModPhys.73.357}. They are based on superconductivity\cite{PhysRev.108.1175}, by incorporating Josephson junctions into superconducting circuits, one can engineer nonlinear quantum oscillators whose discrete energy levels serve as qubit states \cite{JOSEPHSON1962251, clarke1970josephson}.
In this section, we briefly introduce the basic principles of superconductivity and then discuss important superconducting qubit implementations, including the Cooper-pair box (charge qubit)\cite{RN20914}, flux qubit\cite{doi:10.1126/science.285.5430.1036}, and phase qubit\cite{PhysRevLett.89.117901}. These systems demonstrate how charge, flux, and phase can be used to store and control quantum information\cite{RN20919}.
\subsection{Superconductivity Phenomena and the Josephson Effect}
\begin{figure}[b]
    \centering
    \includegraphics[width=\linewidth]{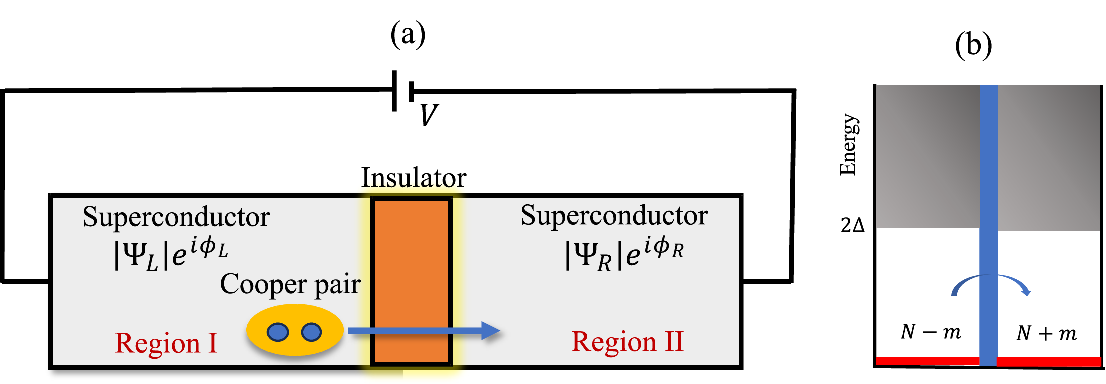}
    \caption{
    (a) Schematic of a Josephson junction consisting of two superconducting electrodes (regions~I and~II) separated by a thin insulating barrier. Cooper pairs coherently tunnel between the two superconductors across the junction.
    (b) Energy-level diagram of a Josephson junction. The superconducting ground state on each side (shown in red) corresponds to all electrons bound into Cooper pairs. This ground state is separated from the continuum of quasiparticle excitations (shown in grey), where electrons are not fully paired, by the superconducting energy gap $2\Delta$. The blue arrow illustrates the coherent tunneling of Cooper pairs between the two superconductors.}
    \label{fig:5}
\end{figure}

The two-level atom discussed above is the simplest model of a quantum system interacting with an external field. Similar two-level systems, called artificial atoms, can be designed in solid-state devices. Unlike natural atoms, these engineered systems offer tunable energy levels, stronger coupling to EM, and easy integration on a chip. They are designed to operate in the microwave regime, making them suitable for modern quantum technologies \cite{PhysRevA.69.062320, you2011atomic, GU20171, Wallraff2004CircuitQED, RevModPhys.93.025005}.
% \subsubsection{Macroscopic Quantum Nature of Superconductivity}
To understand how artificial atoms are designed, we first need to understand the basic properties of superconductivity \cite{PhysRev.108.1175}. Superconductivity is a macroscopic quantum phenomenon that appears in certain materials when they are cooled below a critical temperature $T_c$. Below this temperature, the material shows zero electrical resistance, meaning current can flow without any energy loss. At the same time, it expels magnetic fields from its interior, a phenomenon known as the Meissner effect \cite{meissner1933neuer}.
The origin of superconductivity lies in the pairing of electrons. Normally, electrons repel each other because they have the same charge. However, inside a superconductor, vibrations of the crystal lattice (called phonons) create an effective attractive interaction between electrons. As a result, electrons with opposite momenta and opposite spins pair up to form, known as Cooper pairs \cite{PhysRev.108.1175}. Each Cooper pair behaves like a single particle with total spin 
$S=0$. An important feature of the superconducting state is that all Cooper pairs share a common quantum state, which can be described by a single macroscopic wavefunction,
\begin{equation}
\Psi(\mathbf{r}) = \sqrt{n_s(\mathbf{r})} \, e^{i\varphi(\mathbf{r})},
\end{equation}
where $n_s$ is the density of superconducting Cooper pairs and $\varphi(\mathbf{r})$ is their common quantum phase. Unlike in normal metals, this phase is well defined across the entire superconductor. This collective behavior makes superconductivity a macroscopic quantum phenomenon.
A superconducting material with an even number of electrons has all electrons paired in the lowest-energy state. To break a Cooper pair and create excitations, a minimum energy is required. This energy is called the superconducting energy gap and is equal to $2\Delta$ as shown in Fig. \ref{fig:5}(b).
These unique collective quantum properties allow superconductors to be used to build artificial quantum systems, often called artificial atoms \cite{ RN20918}.
\subsection{Josephson Junction:}
\begin{figure}
    \centering
    \includegraphics[width=\linewidth]{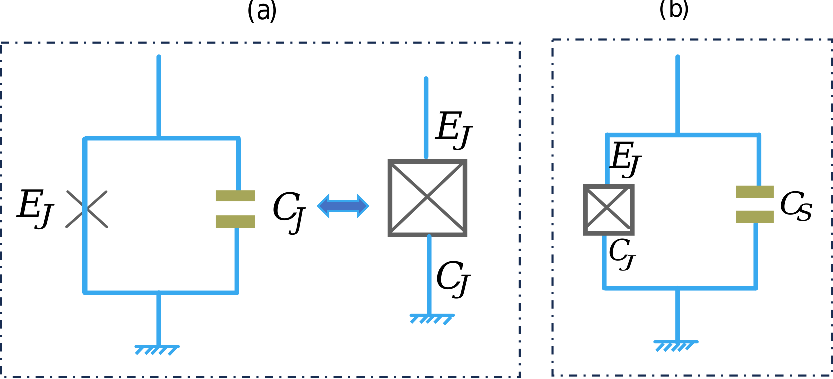}
   \caption{Circuit diagrams. (a) A single Josephson junction with Josephson energy $E_J$ and junction capacitance $C_J$. (b) A Cooper pair box consisting of a Josephson junction ($E_J, C_J$) in parallel with a shunt capacitance $C_S$.}
    \label{fig:6}
\end{figure}

Josephson junctions (JJ) are devices that consists of two superconducting films (electrodes) separated by a thin insulating oxide barrier, across which Cooper pairs can tunnel coherently, as illustrated in Fig.~\ref{fig:5}(a). Although the barrier is insulating, phase coherence allows a dissipationless supercurrent to flow across the junction.
Let $\varphi_L$ and $\varphi_R$ be the macroscopic phases of the two superconductors. The current flowing through the junction depends on the phase difference \cite{tinkham2004bcs}, $\varphi = \varphi_L - \varphi_R$.
The dynamics of the junction are governed by the Josephson relations,
\begin{equation}
I = I_c \sin\varphi, \quad 
\frac{d\varphi}{dt} = \frac{2 \pi}{\Phi_0} V,
\end{equation}
where $\Phi_0 = h/2e$ is the superconducting flux quantum, $I_c$ is the critical current, and $V$ is the voltage across the junction. The first relation corresponds to the DC Josephson effect, where a supercurrent can flow through the junction even in the absence of an applied voltage. The second relation describes the AC Josephson effect, where a constant voltage causes the phase to evolve in time, resulting in an oscillating current. 
Together, these relations show that the Josephson junction behaves like a special type of inductor. Using the inductance definition $V = L_J \, dI/dt$, together with the Josephson current-phase relation, one obtains an effective phase-dependent inductance,
\begin{equation}
L_J(\varphi) = \frac{\Phi_0}{2\pi I_c \cos\varphi}.
\end{equation}
Because this inductance depends on the phase as $\cos \varphi$, this makes the Josephson junction a nonlinear inductive element, which is the key ingredient for building superconducting qubits. In circuit diagrams, it is usually represented by a cross (×) symbol indicating a nonlinear inductor, often accompanied by a parallel capacitance, as shown in Fig. \ref{fig:6}(a),  where the Josephson energy $E_J$ characterizes the nonlinear inductance and $C_J$ represents the junction capacitance.

For a quantum description, we introduce the charge basis $|N\rangle$, where $N$ denotes the number difference of Cooper pairs transferred across the junction. The phase difference $\hat{\varphi}$ and the number operator $\hat{N}$ form a pair of conjugate variables and satisfy the commutation relation, $[\hat{\varphi}, \hat{N}] = i$. In this basis, coherent tunneling of Cooper pairs is described by the Josephson Hamiltonian,
\begin{equation}
H_{\mathrm{JJ}} = -\frac{E_J}{2}\sum_{N=-\infty}^{\infty}
\left(
|N\rangle\langle N+1| + |N+1\rangle\langle N|
\right),
\end{equation}
where $E_J$ is the Josephson energy ~\cite{clarke1970josephson}.
The Hamiltonian $H_{\mathrm{JJ}}$ induces transitions that increase or decrease $N$ by one, corresponding to the tunneling of a single Cooper pair. Formally, it is equivalent to a one-dimensional tight-binding model with nearest-neighbor hopping amplitude $E_J$.
The eigenstates are plane-wave-like superpositions in the charge basis,
\begin{equation}
|\varphi\rangle = \sum_{N=-\infty}^{\infty} e^{-iN\varphi} |N\rangle,
\end{equation}
which leads to the cosine energy-phase relation,
\begin{equation}
H_{\mathrm{JJ}} |\varphi\rangle = -E_J \cos\varphi \, |\varphi\rangle.
\end{equation}
Next, we discuss the Cooper-pair box, a minimal and conceptually clear platform for exploring quantum coherence and controllable two-level dynamics in superconducting circuits\cite{clarke1988quantum, Martinis2020}. 
\subsection{Charge Qubit: The Cooper-Pair Box}
The charge qubit, also known as the Cooper-pair box (CPB), was one of the earliest superconducting devices to exhibit quantum coherence, providing experimental evidence of superpositions between distinct charge states. It also marked the first superconducting circuit in which coherent control and temporal oscillations (Rabi oscillations) were demonstrated \cite{RN20914}.
In an appropriate parameter regime, the CPB behaves as an effective two-level system.
\subsubsection{Charging Effects and Josephson Coupling}
\begin{figure}[t]
    \centering
    \includegraphics[width=\linewidth]{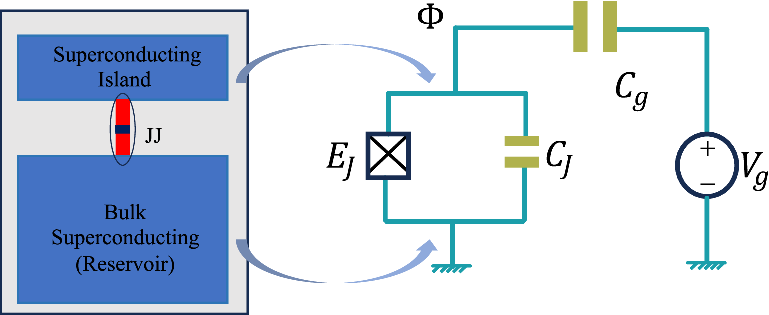}
    \caption{Schematic of a Cooper pair box. A small superconducting island is coupled to a bulk superconducting reservoir via a Josephson junction, modeled by a capacitance $C_J$ in parallel with an effective nonlinear inductor characterized by the Josephson coupling energy $E_J$. The finite charging energy required to add an extra Cooper pair to the island is denoted by $E_C$. A gate electrode, capacitively coupled to the island, applies a gate voltage that induces a dimensionless gate charge $N_g$, which serves as an external control parameter for the system.}
    \label{fig:7}
\end{figure}

The Cooper-pair box (CPB) consists of a small superconducting island connected to a bulk superconducting reservoir through a JJ characterized by capacitance $C_J$ and Josephson energy $E_J$. In addition, the island is capacitively coupled to a gate electrode through a gate capacitance $C_g$, which allows external control of the island charge. The simplest circuit diagram for CPB is shown in Fig. \ref{fig:6}(b) and its connection with reservoir and gate voltage is shown in Fig. \ref{fig:7}.
Because the island is sufficiently small, the addition or removal of a single Cooper pair produces a significant electrostatic energy cost. If $N$ denotes the number of excess Cooper pairs on the island, the total charge is quantized as
\begin{equation}
    Q =  2eN,
\end{equation}
where $e$ is the elementary charge. Because the island is connected to the external circuit only through a single Josephson junction, the Hamiltonian of an isolated CPB can be written as
\begin{equation}
H
= \frac{\hat{Q}^{2}}{2C_J}
- E_J \cos\!\left( \frac{2\pi \hat{\Phi}}{\Phi_0} \right),
\end{equation}
where $\hat{Q}$ is the charge operator on the island, $\hat{\Phi}$ is the magnetic flux across the junction, and $\Phi_0$ is the superconducting flux quantum. The operators $\hat{Q}$ and $\hat{\Phi}$ form a pair of canonically conjugate
variables satisfying the commutation relation
$[\hat{\Phi}, \hat{Q}] = i\hbar.$
It is convenient to express the Hamiltonian in terms of the Cooper-pair number operator $\hat{N}$ and the superconducting phase operator $\hat{\varphi}$,
defined through
\begin{equation}
\hat{Q} = 2e\,\hat{N}, \qquad
\hat{\Phi} = \varphi_0 \hat{\varphi},
\end{equation}
with $\varphi_0 = \Phi_0 / 2\pi$. In these variables, the Hamiltonian takes the
compact form
\begin{equation}
H
= 4E_C \hat{N}^2 - E_J \cos \hat{\varphi},
\label{eq:CPB_basic}
\end{equation}
where
\begin{equation}
E_C = \frac{e^2}{2C_J},
\end{equation}
is the single-electron charging energy, so that a Cooper pair carries an electrostatic energy $4E_C$. The operators $\hat{N}$ and $\hat{\varphi}$ follows the canonical commutation relation $[\hat{\varphi}, \hat{N}] = i.$
In this Hamiltonian, the first term represents the charging energy, which increases when extra Cooper pairs are added to the small superconducting island. The second term represents the Josephson energy, which allows Cooper pairs to tunnel through the junction.
However, an isolated CPB cannot be easily controlled. To tune the qubit, we apply a gate voltage using a nearby electrode that is capacitively coupled to the island, as shown in Fig. \ref{fig:7}. Because the island is very small, even adding or removing a single Cooper pair significantly changes its energy. 
A gate voltage $V_g$ is applied through a gate capacitor $C_g$, it induces a charge 
\begin{equation}
Q_g = C_g V_g,
\end{equation}
on the island. The electrostatic energy then depends on the difference between the actual island charge $Q$ and the induced gate charge $Q_g$.
The Hamiltonian of the gate-controlled CPB takes the form
\begin{equation}
H
= \frac{(\hat{Q} - Q_g)^2}{2C_\Sigma}
- E_J \cos\!\left( \frac{2\pi \hat{\Phi}}{\Phi_0} \right),
\end{equation}
where, $C_\Sigma = C_J + C_g$ is the total capacitance of the island. We now express the Hamiltonian in terms of more convenient variables: the dimensionless Cooper-pair number operator $\hat{N} = \hat{Q}/{2e}
$ and the phase operator 
$\hat{\varphi} = {2\pi \hat{\Phi}}/{\Phi_0}$. In these variables, the Hamiltonian becomes
\begin{equation}
H
= E_C (\hat{N} - N_g)^2 - E_J \cos \hat{\varphi},
\end{equation}
where $E_C = (2e)^2/2C_\Sigma$ is the charging energy and $N_g = Q_g/2e$ is the dimensionless gate charge controlled by the applied voltage. The parameter $N_g$ acts as a tuning knob that allows us to adjust the energy levels of the Cooper-pair box.
By tuning the ratio $E_J/E_C$ and the gate voltage $V_g$, one can control the energy spectrum of the device. This tunability makes the Cooper-pair box a powerful and flexible platform for studying quantum coherence and for implementing qubit operations \cite{RevModPhys.93.025005}.
\subsubsection{Charge-Basis Representation}
To better understand the Cooper-pair box, it is useful to express the Hamiltonian in the charge basis, where the number of Cooper pairs on the island is well defined. Starting from the commutation relation $[\hat{\varphi}, \hat{N}] = i$, one can derive the more general relation
\begin{equation}
    [e^{ik\hat{\varphi}}, \hat{N}] = \sum_{m=0}^\infty \frac{(ik)^m}{m!}[\hat{\varphi}^m, \hat{N}] = -k e^{ik\hat{\varphi}}.
\end{equation}
Setting $k=1$ reduces to $[e^{i\hat{\varphi}}, \hat{N}] = -e^{i \hat{\varphi}}$. 
Let $\ |N\rangle $ denote the charge eigenbasis, defined by $\hat{N}|N\rangle= N|N \rangle$. Using the above commutation relation, one finds that
\begin{equation}
    e^{\pm i \hat{\varphi}}|N\rangle = |N\pm 1 \rangle.
\end{equation}
This shows that the operators $e^{\pm i\hat{\varphi}}$ act as ladder operators in the charge basis, increasing or decreasing the number of Cooper pairs by one. Physically, this corresponds to a single Cooper pair tunneling across the Josephson junction. Using the completeness relation $\sum_m |N \rangle \langle N| = 1$, and expressing the Josephson term as
\begin{equation}
    \cos \hat{\varphi} = \frac{1}{2}\left(e^{i\hat{\varphi}} + e^{-i\hat{\varphi}}\right),
\end{equation}
the Cooper-pair box Hamiltonian can be written in the charge basis as
\begin{equation}
\begin{aligned}
H = & \sum_{N} 4E_C (N - N_g)^2\, |N\rangle\langle N|\\
& - \frac{E_J}{2} \sum_{N}
\left( |N\rangle\langle N+1| + |N+1\rangle\langle N| \right).
\end{aligned}
\end{equation}
This expression clearly shows two important physical effects. The charging term (first line) assigns an energy cost depending on how many Cooper pairs are on the island. The Josephson term (second line) allows coherent tunneling between neighboring charge states. Thus, the system behaves like a particle hopping between discrete charge states. 
\begin{figure}
    \centering
    \includegraphics[width=0.98\linewidth]{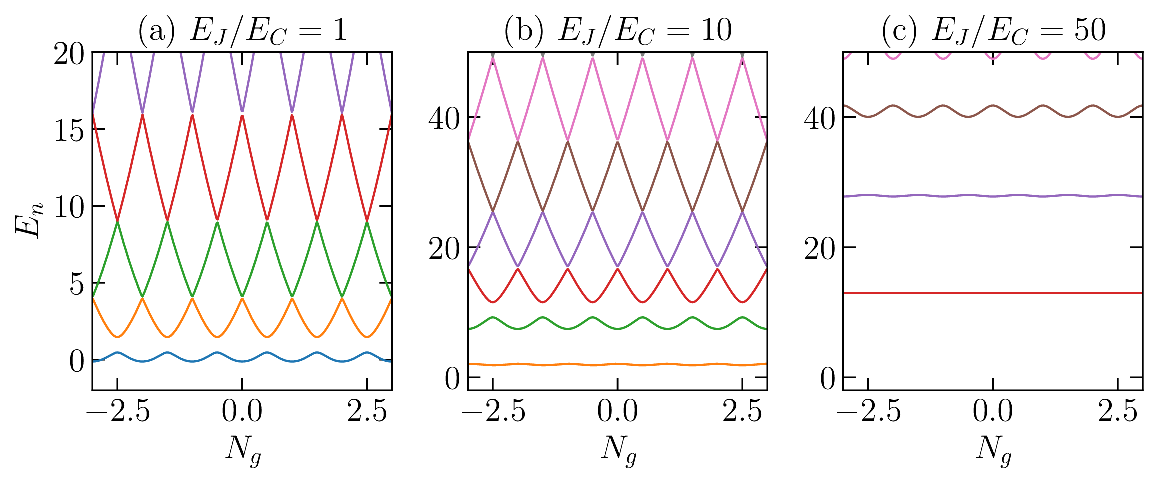}
    \caption{Energy levels $E_n(N_g)$ of the Cooper pair box for three $E_J/E_C$ ratios: (a) $E_J/E_C=1$ (charge qubit regime), (b) $E_J/E_C=10$ (intermediate regime), and (c) $E_J/E_C=50$ (transmon regime).}
\label{fig:8}
\end{figure}
Even though Cooper pairs are discrete charge carriers, $N_g$ can be varied 
continuously by adjusting the applied gate voltage. 
This makes $N_g$ a powerful experimental control knob for tuning 
the energy levels of the device.
By diagonalizing the above charge basis Hamiltonian, 
Fig.~\ref{fig:8} shows the eigenenergies $E_n(N_g)$ versus gate charge $N_g$ at fixed $E_J/E_C$ ratios, revealing distinct quantum regimes. For $E_J/E_C = 1$ (Charge Qubit Regime) \cite{RN20914}, the charging energy and the Josephson tunneling energy are comparable. The energy levels show a strong dependence on the dimensionless gate charge $N_g$ 
and exhibit several anticrossings. 
In this regime, the system is highly sensitive to charge fluctuations and behaves as a charge qubit. For $E_J/E_C = 10$ (intermediate regime), Josephson tunneling becomes stronger, increasing phase coherence. The energy bands become smoother, and sensitivity to charge noise begins to decrease. For $E_J/E_C = 50$ (transmon-like regime), here $E_J \gg E_C$, meaning that charging effects are much weaker 
than the Josephson energy. The energy spectrum becomes nearly parabolic and depends only weakly on the gate charge $N_g$ \cite{PhysRevA.76.042319, PhysRevB.77.180502}. 
In this regime, the device is far less sensitive to charge noise and is therefore ideal for stable qubit operation. This is the basic principle behind the transmon qubit.
\subsubsection{Bosonic Representation and Anharmonicity}
\begin{figure}[ht]
    \centering
    \includegraphics[width=\linewidth]{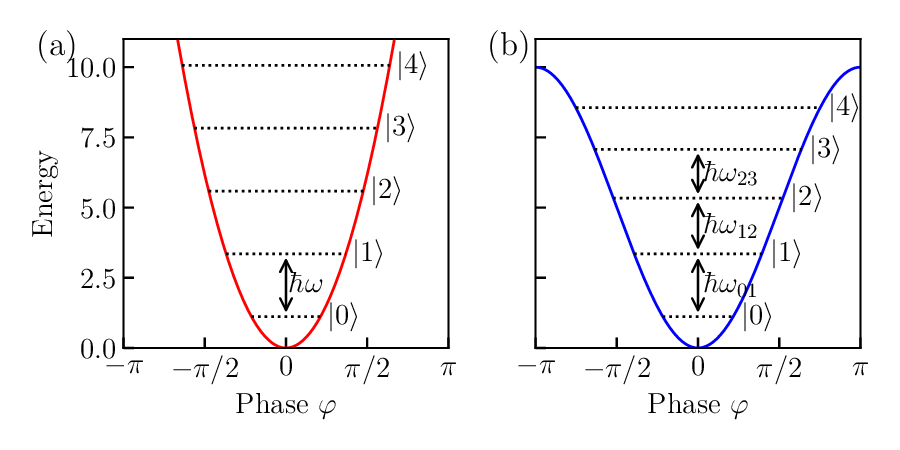}
    \caption{(a) Effective potential energy as a function of the superconducting phase difference $\varphi$ for the linear inductive circuit, exhibiting a purely harmonic spectrum with equally spaced energy levels. (b) Effective potential energy of the Josephson junction circuit, for a nonlinear inductive circuit, showing an anharmonic potential and non-equidistant energy levels, enabling its operation as a two-level system.}
    \label{fig:9}
\end{figure}

In the transmon-like regime ($E_J \gg E_C$), the energy levels depend only weakly on $N_g$, as we saw in Fig.~\ref{fig:8}(c). We can tune $N_g$ to a special point (around $N_g = 1/2$) where charge noise barely affects the system like finding the quietest spot on a bumpy road. Here, the phase $\hat{\varphi}$ jiggles slightly around the bottom of the Josephson potential well $-E_J \cos \hat{\varphi}$, which looks like a wavy washboard. Expand the cosine for small jiggles (like Taylor-expanding $\cos x \approx 1 - x^2/2 + x^4/24$ near $x=0$)
\begin{equation}
- E_J \cos \hat{\varphi}
\simeq
- E_J
+ \frac{E_J}{2}\hat{\varphi}^2
- \frac{E_J}{24}\hat{\varphi}^4
+ \mathcal{O}(\hat{\varphi}^6).
\end{equation}
Ignore the constant $-E_J$. The full Hamiltonian simplifies to
\begin{equation}
\hat{H}
\simeq
4E_C \hat{N}^2
+ \frac{E_J}{2}\hat{\varphi}^2
- \frac{E_J}{24}\hat{\varphi}^4.
\label{eq:duffing_start}
\end{equation}
The first two terms are exactly a quantum harmonic oscillator: charge $\hat{N}$ provides kinetic energy, phase $\hat{\varphi}$ provides potential energy (like a spring). Introducing canonical bosonic ladder operators $\hat{b}$ and $\hat{b}^\dagger$ via
\begin{equation}
\hat{\varphi}
= \varphi_{\rm zpf}(\hat{b}+\hat{b}^\dagger), \qquad
\hat{N}
= i N_{\rm zpf}(\hat{b}^\dagger-\hat{b}),
\end{equation}
with zero-point fluctuations
\begin{equation}
\varphi_{\rm zpf} = \left(\frac{2E_C}{E_J}\right)^{1/4},
\qquad
N_{\rm zpf} = \left(\frac{E_J}{32E_C}\right)^{1/4},
\end{equation}
The harmonic part diagonalizes neatly to
\begin{equation}
\hat{H}_0
= \hbar \omega_q \left(\hat{b}^\dagger\hat{b}+\frac{1}{2}\right),
\end{equation}
with the plasma frequency $\hbar\omega_q = \sqrt{8E_CE_J}$.
The quartic $-\frac{E_J}{24} \hat{\varphi}^4$ in Eq.~(\ref{eq:duffing_start}) term adds nonlinearity. Plugging in the boson expansion and keeping only terms that conserve particle number, it becomes
\begin{equation}
\hat{H}_{\rm nl}
= -\frac{E_J}{24}\varphi_{\rm zpf}^4
(\hat{b}+\hat{b}^\dagger)^4.
\end{equation}
Applying the RWA and retaining only number-conserving terms, we obtain
\begin{equation}
\hat{H}_{\rm nl}
= -\frac{E_C}{2}\hat{b}^\dagger\hat{b}^\dagger\hat{b}\hat{b}.
\end{equation}
This is a Kerr or Duffing nonlinearity \cite{walls2008quantum}. It makes energy levels unevenly spaced (anharmonic), turning the bare oscillator into a qubit. The two plots of a simple harmonic oscillator and adding anharmonicity are shown in Fig.~\ref{fig:9}(a) and Fig.~\ref{fig:9}(b). It beautifully shows the equally spaced (harmonic) vs.\ unequally spaced (anharmonic) energy levels. The full CPB Hamiltonian can be written as
\begin{equation}
\hat{H}_{\rm CPB}
= \hbar \omega_q \hat{b}^\dagger \hat{b}
- \frac{E_C}{2}\hat{b}^\dagger \hat{b}^\dagger \hat{b} \hat{b}.
\label{iii}
\end{equation}
Energy to go from $|0\rangle$ to $|1\rangle$ is $\hbar \omega_q$; to $|2\rangle$ is less than $2\hbar \omega_q$ due to the negative Kerr term perfect for addressing just the qubit transition with microwaves.
This anharmonicity is essential for qubit operation~\cite{JOSEPHSON1962251}. Without the nonlinear term, the system would behave as a purely harmonic oscillator with equally spaced energy levels Fig.~\ref{fig:9}(a), making it impossible to selectively address a single transition using resonant driving. The anharmonicity lifts this degeneracy, creating unequally spaced levels Fig.~\ref{fig:9}(b) that spectrally isolate the qubit transition ($|0\rangle \leftrightarrow |1\rangle$) while suppressing unwanted excitations to higher states. This confines the dynamics to the computational subspace, justifying the two-level qubit approximation while enabling controlled analysis of leakage and nonlinear effects. 
\subsection{Flux qubit}
\begin{figure}[b]
    \centering
    \includegraphics[width=\linewidth]{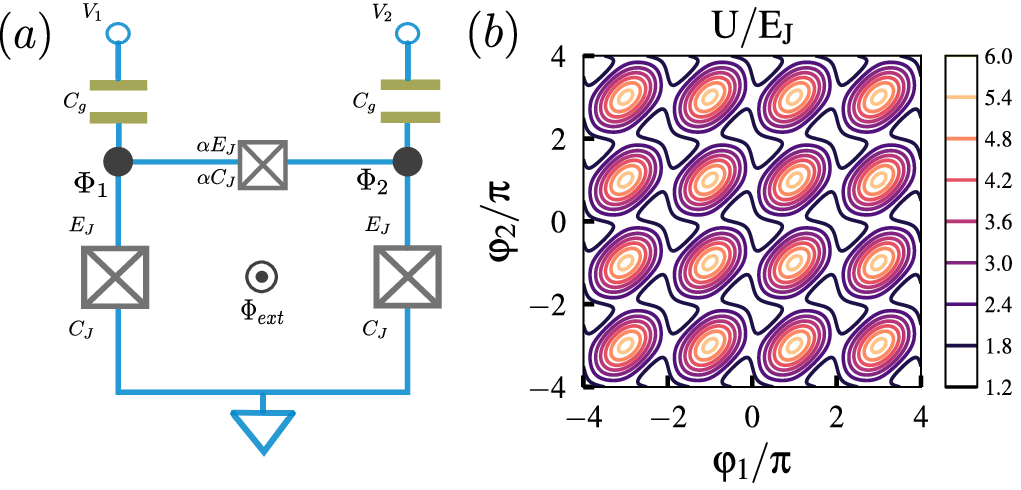}
    \caption{(a) A lumped-element circuit diagram of a flux qubit. (b) The potential energy of the three-junction flux qubit for $\alpha=0.8$ and $k=0.5$.}
    \label{fig:fig8}
\end{figure}
The transmon is at the epicenter of existing state-of-the-art superconducting NISQ devices \cite{Acharya2025, Kim2023}. However, the shunting capacitance in a transmon reduces the anharmonicity in exchange for low decoherence rates. Compared with transmon, flux qubits are versatile platforms that can behave like transmon \cite{Yan2016}, exhibit two-well tunneling hybridization \cite{doi:10.1126/science.1081045}, and exhibit many other intermediate behaviors \cite{10.1063/5.0047690}. In this section, we discuss the three-junction persistent-current flux qubit \cite{PhysRevB.60.15398, doi:10.1126/science.285.5430.1036} as shown in Fig. \ref{fig:fig8} (a). One junction is smaller in area than the other two junctions by a factor $\alpha\sim0.75$ with a typical value of capacitance $\alpha C_J$ and Josephson potential energy $\alpha E_J$. the flux qubit is connected to the gate voltages $V_1$ and $V_2$ via the gate capacitances $C_g\approx\gamma C_J$. We have assumed the symmetric large junctions with Josephson potential energy $E_J$, capacitance $C_J$, and gate capacitances $C_g$ for simplicity. The small junction controls the fluxoid in and out of the loop, and the large junctions add inductance to the loop, which otherwise has negligible inductance. Applying the fluxoid quantization condition
\begin{equation}
\varphi_1-\varphi_2+\varphi_3=-2k\pi,
\end{equation}
where $k=\Phi_{ext}/{\Phi_0}$ and $\Phi_0=h/{2e}$ is the superconducting magnetic flux quantum, the potential energy of the circuit can be expressed as
\begin{equation}
\frac{U}{E_J}=2-\cos\varphi_1-\cos\varphi_2+\alpha\{1-\cos(2k\pi+\varphi_1-\varphi_2)\}.
\end{equation}
The magnetic flux through the smaller junction $\varphi_3$ is written in terms of the magnetic fluxes through the large junctions $\varphi_1,\varphi_2$, and the quantized applied external flux $k=\Phi_{ext}/\Phi_0$ in the loop. The electrostatic energy stored in the capacitor is
\begin{equation}
T=\frac{1}{2}\left(\frac{\Phi_0}{2\pi}\right)^2\dot{\varphi}^TC\dot{\varphi},
\end{equation}
where $\varphi=(\varphi_1\quad\varphi_2)^T$ and
\begin{equation}
C=C_J\begin{pmatrix}
    1+\alpha+\gamma & -\alpha\\
    -\alpha & 1+\alpha+\gamma
\end{pmatrix}.
\end{equation}
We have neglected the constant term $-\frac{1}{2}V_g^TC_gV_g$ with $V_g=(V_1\quad V_2)^T$, and $C_g=\gamma C_J\mathds{1}$, where $\mathds{1}$ represents the $2\times 2$ identity matrix. Considering the Lagrangian $\mathcal{L}=T-U$, the canonically conjugate momenta $P=(P_1\quad P_2)^T=\left(\frac{\Phi_0}{2\pi}\right)^2C\dot{\varphi}$, the Hamiltonian of the flux qubit circuit can be obtained as
\begin{align}
\mathcal{H}=& P^T\dot{\varphi}-\mathcal{L}\nonumber\\
=&\frac{1}{2}\left(\frac{2\pi}{\Phi_0}\right)^2P^TC^{-1}P+U\nonumber\\
=&\frac{1}{2}\left(\frac{2\pi}{\Phi_0}\right)^2\frac{[(1+\alpha+\gamma)(P_1^2+P_2^2)+2\alpha P_1P_2]}{C_J[(1+\alpha+\gamma)^2-\alpha^2]}+U.
\end{align}
The Hamiltonian can be further simplified by considering the notations $P_{\pm}=(P_1\pm P_2)/2$ and $\varphi_{\pm}=(\varphi_1\pm\varphi_2)/2$. The final expression of the Hamiltonian can be expressed as
\begin{align}
\mathcal{H}=&\frac{1}{2}\left(\frac{P_+^2}{m_+}+\frac{P_-^2}{m_-}\right)+\nonumber\\
&E_J[2+\alpha-2\cos\varphi_+\cos\varphi_--\alpha\cos(2k\pi+2\varphi_-)]
\end{align}
where $m_+=\left(\frac{\Phi_0}{2\pi}\right)^2\frac{C_J}{2}(1+\gamma)$ and $m_-=\left(\frac{\Phi_0}{2\pi}\right)^2\frac{C_J}{2}(1+2\alpha+\gamma)$ can be considered as the effective masses, while $P_{\pm}=-i\hbar\frac{\partial}{\partial\varphi_{\pm}}$ can be considered as the reduced momenta. The potential energy can be assumed as a two-dimensional double-well profile for an externally applied flux quantum $k\approx0.5$, as shown in Fig. \ref{fig:fig8} (b). The states in each well correspond to clockwise or counterclockwise circulating persistent currents with magnitude $I_p$ around the loop. The lowest-energy circulating-current states in each well form the two-level flux qubit. These states have energy,
\begin{equation}
\pm\frac{\hbar\mathcal{E}}{2}=\pm I_p\Phi_0(k-1/2),
\end{equation}
where, at $k=1/2$ the lowest two-level states are degenerate. Quantum tunneling through the double-well barrier hybridizes the two states, opening an avoided crossing of strength $\Delta$. Therefore, the effective two-level system Hamiltonian for the flux qubit is
\begin{equation}
\mathcal{H}_{TLS}^F=-\frac{1}{2}[2I_p\Phi_0(k-1/2)\sigma_x+\hbar\Delta\sigma_z],
\end{equation}
where, $\sigma_{x,z}$ are Pauli spin matrices.

\subsection{Phase qubit}
\begin{figure}[b]
    \centering
    \includegraphics[width=\linewidth]{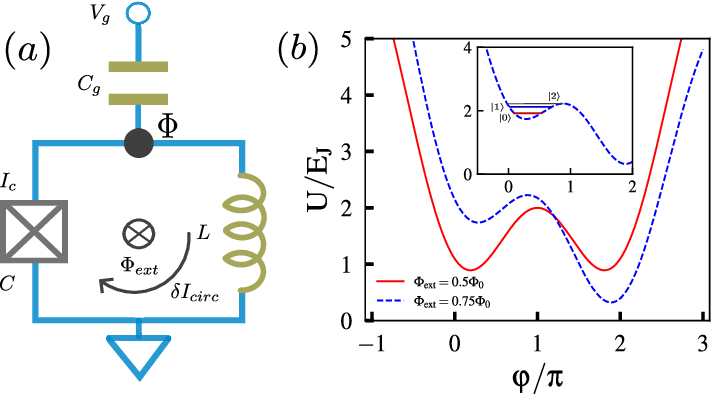}
    \caption{(a) A lumped-element circuit diagram of a Phase qubit. (b) Potential energy of an RF SQUID for different values of $\Phi_{ext}$ and $\beta_L=4.5$. The inset resembles the washboard-type potential characteristic of a current-biased phase qubit.}
    \label{fig:fig9}
\end{figure}
The earliest observations of macroscopic quantum-mechanical behaviour were made in current-biased phase qubits \cite{PhysRevLett.89.117901, Martinis2020}. The phase qubit is built by current-biasing a JJ close to its critical current $I_c$ as shown in Fig. \ref{fig:fig9} (a). The junction is placed in a loop of inductance $L$ to isolate it from its environment, forming an RF SQUID \cite{PhysRevLett.93.180401}. The qubit loop can be biased by an external magnetic flux $\Phi_{ext}$, effectively biasing the junction with the circulating current $\delta I_{circ}$ in the loop. In this condition, a washboard-type potential forms, accommodating a few quantized energy levels. The potential energy of the phase qubit can be expressed as 
\begin{equation}
U=E_J\left[1-\cos\varphi+\frac{(\varphi-2\pi\Phi_{ext}/\Phi_0)^2}{2\beta_L}\right]
\end{equation}
where, $\beta_L\equiv2\pi LI_c/\Phi_0$. Considering the electrostatic energy stored in the capacitor $T=C\dot{\Phi}^2/2$, and the Lagrangian $\mathcal{L}=T-U$, the Hamiltonian of the phase qubit loop can be obtained as
\begin{equation}
\mathcal{H}=\frac{p^2}{2m}+U
\end{equation}
where $p=-i\hbar\frac{\partial}{\partial\varphi}$ and $m=C\left(\frac{\Phi_0}{2\pi}\right)^2$. The RF SQUID can be operated as a phase qubit for $1<\beta_L<4.6$ with a given value of $I_c$ \cite{, lisenfeld_2025_j9e04-01y12}. By varying the external flux $\Phi_{ext}$, one can realize a tilted double-well potential with a potential barrier height $\Delta U$. The potential barrier can be reduced by tuning he external flux as shown in Fig. \ref{fig:fig9} (b). The shallow potential well becomes anharmonic with distinct $\omega_{01}$ and $\omega_{12}$ transitions for the lowest energy-level states $|0\rangle,|1\rangle$ and $|2\rangle$. The excited energy-level state $|2\rangle$ tunnels rapidly into the relatively deeper potential well. Driving the $\omega_{12}$ transition of the qubit results in tunneling into the deep well. The tunneling event is read out by a DC SQUID to look for a change in flux ($\approx\Phi_0$) or junction voltage, indicating the qubit was in state $|1\rangle$, whereas no change in flux indicates state $|0\rangle$. The two-level system model Hamiltonian of the phase qubit (RF SQUID) can be approximated as
\begin{equation}
\mathcal{H}_{TLS}^{P}=-\frac{1}{2}\left[\hbar\omega_{01}\sigma_z+\sqrt{\frac{\hbar}{2\omega_{01}C}}\delta I_{circ}(\sigma_x+\chi\sigma_z)\right],
\end{equation}
where, $\chi=\sqrt{\hbar\omega_{01}/3\Delta U}$ and $\sigma_{x,y,z}$ are the Pauli spin matrices. Many advanced superconducting qubit architectures derived from the basic charge qubit, flux qubit, and phase qubit are reviewed in recent articles and lecture notes \cite{Oliver2013superconducting, Kockum2019, 10.1063/1.5089550}.
\section{Multilevel circuit QED}\label{Sec:IV}
In this section, we shift our focus from two-level systems (qubits) to multilevel systems. Multi-state quantum coherence has been demonstrated in SQCs using single anharmonic JJ based resonators \cite{PhysRevLett.103.193601, PhysRevLett.104.193601}, coupled qubit-resonator \cite{PhysRevLett.102.243602} and multi-qubit systems \cite{PhysRevLett.103.150503}. A three-level system (qutrit) is the simplest multi-level system. Generally, natural three-level systems are classified into $\Lambda$, $\Xi$, and $V$ configurations allowed by the parity selection rules. However, the parities of the eigenstates of SQCs can be engineered by external parameters, such as magnetic fluxes and electric fields. In a flux qubit circuit, the external magnetic flux can only induce the transitions between states with different parities for $k=0.5$. However, transition between any two states is possible when $k\neq 0.5$ \cite{PhysRevLett.95.087001}. Such a scenario allows the realization of a $\Delta$-type configuration that does not naturally exist. Many quantum optical phenomena have been demonstrated with superconducting qutrits. The engineered superconducting $\Delta$ qutrits enable new applications such as single-photon generation \cite{PhysRevB.75.104516}, qubit cooling \cite{PhysRevLett.100.047001}, downconversion \cite{PhysRevB.76.205416}, and microwave lasing \cite{PhysRevLett.115.223603}, among others. This section discusses the implementation of coherent control of coherence and population in a dressed-state-engineered $\Delta$ system in circuit QED.
\subsection{Jaynes-Cummings model in circuit QED system}
\begin{figure}[b]
    \centering
    \includegraphics[width=0.8\linewidth]{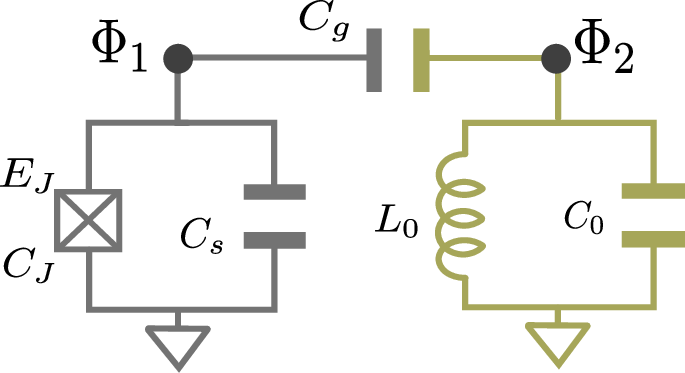}
    \caption{A lumped-element circuit diagram of a transmon capacitively coupled to an LC oscillator.}
    \label{fig:fig10}
\end{figure}

 In analogy with CQED \cite{Haroche2020}, this section discusses the interaction of a quantum harmonic LC oscillator with a transmon as shown in Fig. \ref{fig:fig10}. The transmon is capacitively coupled to a microwave LC oscillator \cite{PhysRevA.69.062320}. The Lagrangian of this circuit can be expressed as
 
\begin{align}
    \mathcal{L}=&\frac{1}{2}(C_J+C_s)\dot{\Phi}_1^2+E_J\cos{\varphi_1}+\frac{1}{2}C_g(\dot{\Phi}_1-\dot{\Phi}_2)^2\nonumber\\
    &+\frac{1}{2}C_0\dot{\Phi}_2^2-\frac{\Phi_2^2}{2L_0}\nonumber\\
    =&\frac{1}{2}\dot{\Phi}^TC\dot{\Phi}+E_J\cos{\varphi_1}-\frac{\Phi_2^2}{2L_0},
\end{align}
where $\Phi=(\Phi_1\quad\Phi_2)^T$ denotes the node fluxes, and
\begin{equation}
C=\begin{pmatrix}
    C_J+c_s+C_g & -C_g\\
    -C_g & C_0+C_g
\end{pmatrix}.
\end{equation}
The canonically conjugate variables of $\Phi_1$ and $\Phi_2$ can be represented as the charges $Q=(Q_1\quad Q_2)^T=C\dot{\Phi}$. Therefore, we can derive the Hamiltonian of the transmon-cavity system by following the Legendre transform
\begin{align}
\mathcal{H}=&Q^T\dot{\Phi}-\mathcal{L}\nonumber\\
=&\frac{1}{2}Q^TC^{-1}Q-E_J\cos{\varphi_1}+\frac{\Phi_2^2}{2L_0}\nonumber\\
=&\frac{Q_1^2}{2C_1}+\frac{Q_2^2}{2C_2}+\frac{C_gQ_1Q_2}{\bar{C}^2}\nonumber\\
&-E_J\cos{\varphi_1}+\frac{\Phi_2^2}{2L_0},
\end{align}
where $\bar{C}^2=(C_gC_\Sigma+C_gC_0+C_\Sigma C_0)$ with $C_\Sigma=C_J+C_s$, $C_1=C_\Sigma+\frac{C_0C_g}{C_0+C_g}$, and $C_2=C_0+\frac{C_\Sigma C_g}{C_\Sigma+C_g}$. The total Hamiltonian of the system can be rearranged as
\begin{align}
\mathcal{H}=&\left(\frac{Q_2^2}{2C_2}+\frac{\Phi_2^2}{2L_0}\right)+\left(\frac{Q_1^2}{2C_1}-E_J\cos{\varphi_1}\right)\nonumber\\
&+\frac{C_gQ_1Q_2}{\bar{C}^2}.
\end{align}
The first two terms inside the parentheses represent the effective Hamiltonian of the harmonic LC oscillator, the third and fourth terms inside another parentheses constitute the effective Hamiltonian of the transmon, and the last term denotes the interaction between the transmon and the oscillator. Following the quantization of node fluxes and charges, we can derive the quantized Hamiltonian

\begin{subequations}

\begin{align}
    \hat{H} & \approx \hat{H}_{c} + \hat{H}_{q} + \hat{H}_I\label{i}\\
    \hat{H}_c & = \hbar \omega_c (\hat{a}^\dagger \hat{a} + 1/2)\label{ii}\\
    \hat{H}_q & = \hbar \omega_q \hat{b}^\dagger \hat{b} - \frac{E_c}{2} \hat{b}^\dagger \hat{b}^\dagger \hat{b} \hat{b}\label{iii}\\
    \hat{H}_I & = -\hbar g (\hat{b}^\dagger - \hat{b}) (\hat{a}^\dagger - \hat{a})\label{iv}
\end{align}
\end{subequations}
where $\hat{a}$ ($\hat{b}$) and $\hat{a}^\dagger$ ($\hat{b}^\dagger$) are the annihilation and the creation operators of the LC oscillator (transmon), respectively. The resonant frequency of the LC oscillator cavity and the transmon (qubit) are  $\omega_c=1/\sqrt{L_0C_2}$ and $\omega_q=\sqrt{8E_CE_J}-E_C$ with the effective charging energy $E_C=e^2/2C_1$, respectively. The interaction strength between the cavity and the transmon is proportional to the product of zero-point charge fluctuations of the cavity and the transmon, which depends on the fine-structure constant $\alpha=\frac{e^2}{4\pi\varepsilon_0\hbar c}\approx\frac{1}{137}$. For a smaller coupling constant $g\ll \omega_r,\omega_q$, under the rotating wave approximation Eq. (\ref{iv}) reduces to
\begin{equation}
    \hat{H}_I = \hbar g (\hat{b}^\dagger \hat{a} + \hat{b} \hat{a}^\dagger). 
\end{equation}
We now project the transmon Hamiltonian onto its two lowest energy levels, denoted by $|g\rangle$ and $|e\rangle$. By replacing the bosonic creation (annihilation) operator $\hat{b}^\dagger(\hat{b})$ with the corresponding spin raising (lowering) operators  $\hat{\sigma}_+(\hat{\sigma}_-)$, we obtain the celebrated Jaynes-cummings Hamiltonian (JCM)
\begin{equation}\label{X}
    \hat{H} = \hbar \omega_r \hat{a}^\dagger \hat{a} +\hbar \frac{\omega_q}{2} \hat{\sigma}_z + \hbar g (\hat{\sigma}_+ \hat{a} + \hat{\sigma}_- \hat{a}^\dagger). 
\end{equation}
where the spinor operators $\hat{\sigma}_z=|e\rangle\langle e|-|g\rangle\langle g|$, $\hat{\sigma}_+=|e\rangle\langle g|$, $\hat{\sigma}_{-}=|g\rangle\langle e|$. We neglect the anharmonic term $- \frac{E_c}{2} \hat{b}^\dagger \hat{b}^\dagger \hat{b} \hat{b}$, which is characteristic of the transmon for a strong coupling with the cavity $g>\kappa$, where $\kappa$ is the cavity decay rate. The zero-point energy $\omega_r/2 $ is also dropped from the total Hamiltonian by following the usual convention. 

\subsection{Dressed state engineering in the Jaynes-Cummings model}\label{DS}

\begin{figure}[ht]
    \centering
    \includegraphics[width=0.9\linewidth]{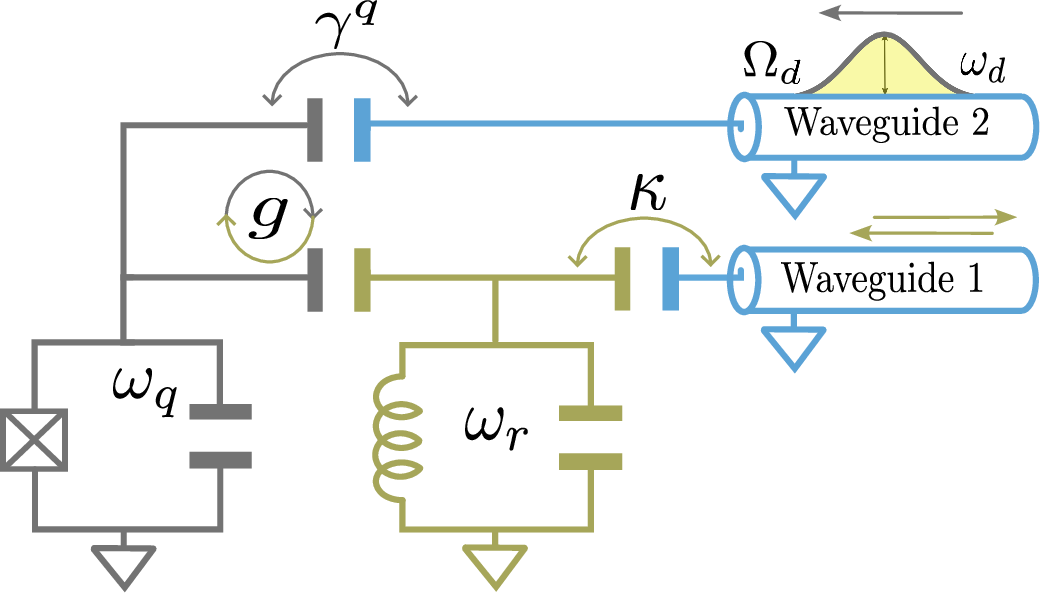}
    \caption{A lumped-element circuit diagram of a transmon capacitively coupled to an LC oscillator with a classical drive driving the transmon with a driving frequency $\omega_d$ and Rabi frequency $\Omega_d$..}
    \label{fig:fig11}
\end{figure}

The dynamics of the JCM Hamiltonian can be solved by finding its stationary states. Considering the Hamiltonian in Eq. (\ref{X}), the interaction term allows only the following transitions
\begin{subequations}
\begin{align}
    |e, n\rangle & \leftrightarrow |g, n+1\rangle,\label{v}\\
    |e, n-1\rangle & \leftrightarrow |g, n\rangle,\label{vi}
\end{align}
\end{subequations}
where the bare states $|\alpha, n\rangle$ span the Hilbert space $|\alpha\rangle\otimes|n\rangle$ of the JCM Hamiltonian, where $\alpha=e,g$, and $n=0,1,2,...$ denote the excitation states of the atom and the resonator, respectively. Using the states $|e,n\rangle$ and $|g,n+1\rangle$ as the basis, we obtain the matrix representation of the JCM Hamiltonian in the $2\times2$ subspace 
\begin{equation}\label{Y}
\hat{H}_n=\hbar
\begin{bmatrix}
    n\omega_r+\frac{1}{2}\omega_q & g\sqrt{n+1}\\
    & \\
    g\sqrt{n+1} & (n+1)\omega_r-\frac{1}{2}\omega_q
\end{bmatrix}.
\end{equation}
The eigenvalues of $\hat{H}_n$ are as follows:
\begin{subequations}
\begin{align}
    E_\pm & = (n+\frac{1}{2})\hbar\omega_r\pm\frac{1}{2}\hbar\Omega(\Delta)\label{vii},\\
    \Omega_n(\Delta) & = [\Delta^2 + 4 g^2 (n+1)]^{1/2},\label{viii}
\end{align}
\end{subequations}
\noindent where $\Omega_n(\Delta)$ with $\Delta=\omega_q-\omega_r$ is the detuning dependent Rabi frequency. The corresponding eigenstates $|n,\pm\rangle$ associated with the eigenvalues $E_\pm$ are given by
\begin{subequations}\label{B}
\begin{align}
    |n,+\rangle & = \cos{\frac{\theta_n}{2}} |e,n\rangle + \sin{\frac{\theta_n}{2}} |g, n+1\rangle\label{vii},\\
    |n,-\rangle & = -\sin{\frac{\theta_n}{2}} |e, n\rangle +\cos{\frac{\theta_n}{2}} |g, n+1\rangle\label{viii},\\
    \noindent  \text{where}\quad\theta_n & = \tan^{-1}\left(\frac{\Omega_n(0)}{\Delta}\right)\label{ix}
\end{align}
\end{subequations}
These stationary states $|n,\pm\rangle$ are known as ``dressed states" or as the Jaynes-Cummings doublets \cite{doi:https://doi.org/10.1002/9783527617197.ch6}. External drive terms added to the JCM Hamiltonian can be used to engineer doubly dressed polariton states. An impedance-matched $\Lambda$ three-level system can be realized in the doubly-dressed polariton basis \cite{PhysRevLett.111.153601}. The model Hamiltonian of a driven cavity/circuit QED system, as shown in Fig. \ref{fig:fig11}, can be expressed as
\begin{equation}\label{Z}
\begin{aligned}
    \hat{H_S} & = \hbar \omega_r \hat{a}^\dagger \hat{a} +\hbar \frac{\omega_q}{2} \hat{\sigma}_z + \hbar g (\hat{\sigma}_+ \hat{a} + \hat{\sigma}_- \hat{a}^\dagger)\\
     & + \hbar \Omega_d [\hat{\sigma}_- e^{i\omega_dt}+\hat{\sigma}_+ e^{-i\omega_dt}],
\end{aligned}
\end{equation}
where an external classical driving field in the second line is added to the JCM Hamiltonian given in the first line of Eq. (\ref{Z}). The term $\Omega_d$ denotes the coupling strength or Rabi frequency of the driving field interacting with the qubit with frequency $\omega_d$. Using a unitary operator $ U=e^{[-i\omega_d(\sigma_z/2+a^\dag a)t]}$, in a frame rotating with frequency $\omega_d$, We obtain the effective Hamiltonian
\begin{equation}\label{A}
\begin{aligned}
    \hat{H_S} & = \hbar \tilde{\omega}_r \hat{a}^\dagger \hat{a} +\hbar \frac{\tilde{\omega}_q}{2} \hat{\sigma}_z + \hbar g (\hat{\sigma}_+ \hat{a} + \hat{\sigma}_- \hat{a}^\dagger)\\
     & + \hbar \Omega_d [\hat{\sigma}_-+\hat{\sigma}_+],
\end{aligned}
\end{equation}
where $\tilde{\omega}_r=\omega_r-\omega_d$ and $\tilde{\omega}_q=\omega_q-\omega_d$ are referred to as cavity-detuning and qubit-detuning, respectively. In the absence of the external driving field ($\Omega_d=0$), the dynamics of the system can be described by the dressed states given in Eq. (\ref{B}). However, in the dispersive coupling regime ({\it i.e.,} $g\ll \Delta$), $\cos\theta_n/2\approx1$, $\sin\theta_n/2\approx0$, and the dressed states take the following form
\begin{equation}\label{C}
    |n-1,+\rangle \approx  |g, n\rangle, \quad |n, -\rangle \approx |e, n\rangle.
\end{equation}
The eigenenergies associated with these states can be expressed as 
\begin{subequations}\label{D}
\begin{align}
    \omega_{g, n} \equiv & \omega_{n-1, +} \approx n(\tilde{\omega}_r+\chi)+{\Delta}/2 \label{xi},\\
    \omega_{e, n} \equiv & \omega_{n, -} \approx n(\tilde{\omega}_r-\chi)+(\tilde{\omega}_q-\chi)+{\Delta}/2 \label{xii},
\end{align}
\end{subequations}

\begin{figure}[t]
    \centering
    \includegraphics[width=\linewidth]{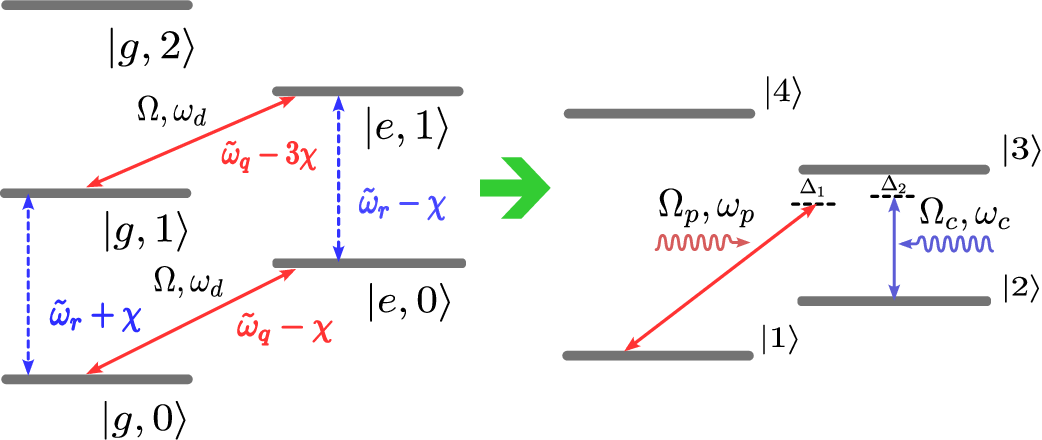}
    \caption{The dressed states in the dispersive regime $|g,0\rangle,|e,0\rangle,|g,1\rangle$, and $|e,1\rangle$ are further mixed by the external field driving the transmon qubit. The states $|g,n\rangle$ and $|e,n\rangle$ (where $n=1,2$) are mixed by the external drive to form the doubly-dressed polariton states $|1\rangle,|2\rangle,|3\rangle$, and $|4\rangle$. The lowest energy-level transitions $|1\rangle\leftrightarrow|3\rangle$ and $|2\rangle\leftrightarrow|3\rangle$ can be driven by a probe and a control field to study EIT in a $\Lambda$ system in circuit QED.}
    \label{fig:fig12}
\end{figure}
where $\chi = g^2/\Delta$ is the dispersive frequency shift. The external classical driving field frequency $\omega_d$ must satisfy the condition $\omega_q-3\chi<\omega_d<\omega_q-\chi$ to operate the driven circuit QED system in the so-called nesting regime ({\it i.e.,} $\omega_{g,0}<\omega_{e,0}<\omega_{e,1}<\omega_{g,1}$). The driving field with coupling strength $\Omega_d$ and frequency $\omega_d$ further dresses the states $|g,0\rangle$ and $|0, -\rangle\approx |e, 0\rangle$ which result in two doubly-dressed polariton states
\begin{subequations}\label{E}
\begin{align}
    |1\rangle & = -\sin{\frac{\theta_l}{2}}|e,0\rangle+\cos{\frac{\theta_l}{2}}|g,0\rangle,\label{xiii}\\
    |2\rangle & = \cos{\frac{\theta_l}{2}}|e,0\rangle+\sin{\frac{\theta_l}{2}}|g,0\rangle,\label{xiii}\\
    \text{where}\quad\tan{\theta_l} & =\frac{2\Omega_d}{(\tilde{\omega}_q-\chi)}.\label{xiv}
\end{align}
\end{subequations}
The transition frequency ($\omega_{ij}=\omega_i-\omega_j$) between the states $|1\rangle$ and $|2\rangle$ is given by 
\begin{equation}\label{F}
    \omega_{21}=\sqrt{(\tilde{\omega}_q-\chi)^2+4\Omega_d^2}.
\end{equation}
Mixing the states $|0, +\rangle\approx|g, 1\rangle$ and $|1, -\rangle\approx|e, 1\rangle$ by the driving field gives us 
\begin{subequations}\label{G}
\begin{align}
    |3\rangle & = -\sin{\frac{\theta_u}{2}}|g,1\rangle+\cos{\frac{\theta_u}{2}}|e,1\rangle,\label{xv}\\
    |4\rangle & = \cos{\frac{\theta_u}{2}}|g,1\rangle+\sin{\frac{\theta_u}{2}}|e,1\rangle,\label{xvi}\\
    \text{where}\quad\tan{\theta_u} & =\frac{2\Omega_d}{(-\tilde{\omega}_q+3\chi)}.\label{xvii}
\end{align}
\end{subequations}
The corresponding energy splitting between the states $|3\rangle$ and $|4\rangle$ is given by
\begin{equation}\label{H}
    \omega_{43}=\sqrt{(\tilde{\omega}_q-3\chi)^2+4\Omega_d^2}.
\end{equation}
The engineered doubly-dressed polariton states are illustrated in Fig. \ref{fig:fig12}. We note that $\omega_{21}=0$ when $\omega_d=\omega_q-\chi$, and $\omega_{43}=0$ when $\omega_d=\omega_q-3\chi$. Therefore, these degeneracies are lifted when the drive field frequency is in the nesting regime $\omega_q-3\chi<\omega_d<\omega_q-\chi$. Since the doubly-dressed polariton state $|3\rangle$ (or $|4\rangle$) is obtained by mixing the states $|g,1\rangle$ and $|e,1\rangle$, externally driving the cavity with two classical fields induces the transitions $|3\rangle\leftrightarrow|2\rangle$ and $|3\rangle\leftrightarrow|1\rangle$. Therefore, it is possible to construct a $\Lambda$ three-level transition with the driven circuit QED system.

\subsection{EIT and ATS in the doubly-dressed polariton states}
The EIT effect can be studied with a $\Lambda$ type atom interacting with two classical driving fields \cite{PhysRevLett.66.2593, RevModPhys.77.633}. A weak resonant probe field interacting with the atom becomes transparent in the medium in the presence of a strong control field. However, the realization of EIT in SQCs is difficult due to the lack of $\Lambda$ systems with suitable metastable states. Following the discussion in the previous section, we introduce a strong external control field with frequency $\omega_c$ to link the transition between the doubly-dressed polariton states $|3\rangle$ and $|2\rangle$. Simultaneously, a weak probe field with frequency $\omega_p$ induces the transition between the states $|3\rangle$ and $|1\rangle$ as demonstrated in recent studies \cite{PhysRevA.93.063827, PhysRevLett.120.083602}. The interaction of the $\Lambda$ system with the control and probe fields can be described by an interaction-picture Hamiltonian
\begin{equation}\label{I}
    \hat{H}_{int}= -\frac{\hbar}{2}[\Omega_p|3\rangle\langle 1|e^{-i\Delta_1t}+\Omega_c|3\rangle\langle 2|e^{-i\Delta_2t}+H.c.]
\end{equation}
where $\Delta_1=\omega_{31}-\omega_p$ and $\Delta_2=\omega_{32}-\omega_c$ are referred to as probe and control field detunings, respectively. The Lindblad master equation for the three-level $\Lambda$ system is given by
\begin{equation}\label{J}
    \dot{\hat{\rho}}=\frac{1}{i\hbar}[\hat{H}_{int},\hat{\rho}] + \sum\limits_{j=1}^{3}\mathcal{L}(\hat{\mathcal{O}}_j)\hat{\rho},
\end{equation}
where $\mathcal{L}(\hat{\mathcal{O}}_j)\rho = (2\hat{\mathcal{O}}_j\hat{\rho}\hat{\mathcal{O}}_j^\dagger-\hat{\rho}\hat{\mathcal{O}}_j^\dagger\hat{\mathcal{O}}_j-\hat{\mathcal{O}}_j^\dagger\hat{\mathcal{O}}_j\hat{\rho})/2$ represents the Lindblad superoperator. Here, the operators $\hat{\mathcal{O}}_j$ denote the jump operators $\hat{\mathcal{O}}_1 = \sqrt{\gamma_{31}}|1\rangle\langle 3|$, $\hat{\mathcal{O}}_2 = \sqrt{\gamma_{32}}|2\rangle\langle 3|$, and $\hat{\mathcal{O}}_3 = \sqrt{\gamma_{21}}|1\rangle\langle 2|$. Perturbatively solving the master equation up to first order of the strength of the probe field, one can obtain the linear susceptibility of the probe field $\chi^{(1)}(-\omega_p,\omega_p)\propto\rho_{31}$. The steady-state population of the three-level system is assumed to be almost in the ground state  ({\it i.e.,} $\rho_{11}\approx1$). The final solution of the linear susceptibility is given by
\begin{equation}\label{K}
    \chi^{(1)}(-\omega_p,\omega_p) = \frac{\delta-\frac{i\gamma_{21}}{2}}{(\delta-\frac{i\Gamma_{31}}{2})(\delta+\Delta_2-\frac{i\gamma_{21}}{2})-\frac{\Omega_c^2}{4}}
\end{equation}
where $\delta=\Delta_1-\Delta_2$ is the two-photon detuning, and $\Gamma_{31}=\gamma_{31}+\gamma_{32}$ denotes the total decay rate of the state $|3\rangle$. The absorption of the probe field is characterized by the imaginary part of the linear susceptibility. Assuming the control field detuning $\Delta_2=0$ for simplicity, we can write
\begin{equation}\label{L}
    \chi^{(1)}(-\omega_p,\omega_p) = \left[\frac{\chi_+}{\delta-\delta_+}-\frac{\chi_-}{\delta-\delta_-}\right],
\end{equation}
where $\chi_{\pm}=\pm(\delta_{\pm}-i\gamma_{21}/2)/(\delta_+-\delta_-)$, and the poles of the denominator are given by 
\begin{equation}\label{M}
    \delta_{\pm} = i\frac{\gamma_{21}+\Gamma_{31}}{4}\pm\frac{1}{2}\sqrt{\Omega_c^2-\frac{1}{4}(\Gamma_{31}-\gamma_{21})^2}.
\end{equation}

\begin{figure}[t]
    \centering
    \includegraphics[width=\linewidth]{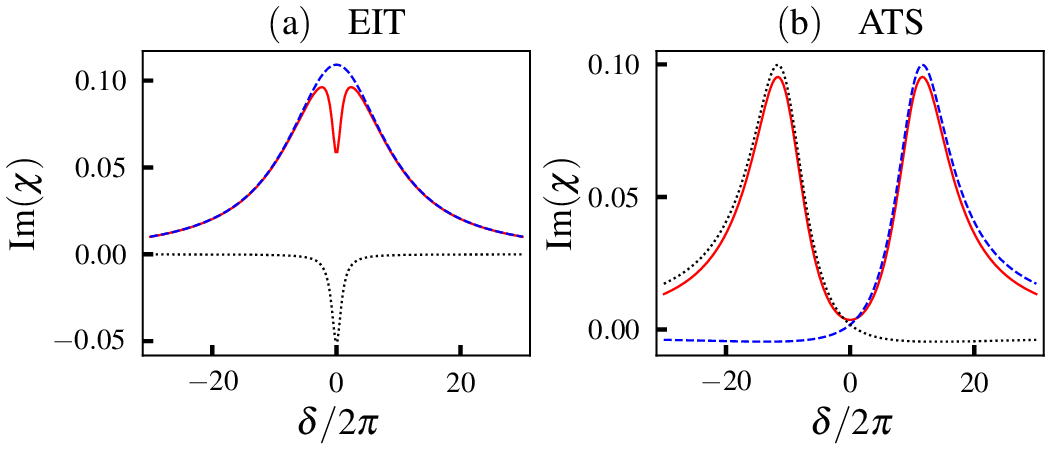}
    \caption{The imaginary part of susceptibility $Im(\chi)$ vs the two-photon detuning $\delta/2\pi$ when (a) the EIT and (b) the ATS conditions are satisfied. }
    \label{fig:fig13}
\end{figure}

For a strong control field $\Omega_c>\frac{1}{2}(\Gamma_{31}-\gamma_{21})$, ATS occurs. In this case, the spectrum of $Im[\chi]$ decomposes into two positive Lorentzians of equal linewidths separated by a distance proportional to $\Omega_c$. The EIT occurs for $\Omega_c<\frac{1}{2}(\Gamma_{31}-\gamma_{21})$, which results from the composition of one broad positive Lorentzian and a narrow negative Lorentzian, both centered at $\delta=0$ as shown in Fig. \ref{fig:fig13}. Both ATS and EIT reduce the absorption of the probe field via different physical mechanisms. The interfering pathways are responsible for the transparency in EIT, whereas splitting of energy levels with a stronger control field causes the ATS \cite{PhysRevLett.107.163604, PhysRevA.93.053838}. EIT has also been studied with several other superconducting circuits \cite{Novikov2016, PhysRevA.103.023710, PhysRevA.102.053721}. The dispersion and group velocities of a probe field in an EIT medium can be controlled by a control field for application in the storage and retrieval of optical or microwave pulse signals \cite{PhysRevLett.109.253603, PhysRevResearch.5.033192, PhysRevResearch.7.L012015}. Recent studies have explored broadband quantum memory in multi-resonator systems based on the principle of photon echo \cite{PhysRevA.95.012338, PhysRevLett.127.010503, PhysRevApplied.19.034011, m9qc-ppk3}

\subsection{Stimulated Raman adiabatic passage}\label{ST}
\begin{figure}[b]
    \centering
    \includegraphics[width=0.8\linewidth]{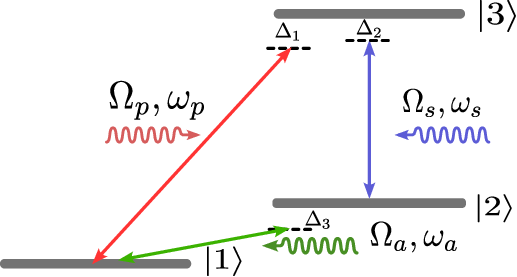}
    \caption{The energy-level transitions $|1\rangle\leftrightarrow|3\rangle$, $|2\rangle\leftrightarrow|3\rangle$, and $|1\rangle\leftrightarrow|2\rangle$ are driven by the pump field, Stokes field, and the counterdiabatic drive field, respectively. STIRAP is studied in the absence of the counterdiabatic drive, whereas saSTIRAP is studied in its presence.}
    \label{fig:fig14}
\end{figure}

The coherent transfer of population between two discrete quantum states of a system is essential for quantum technology. The STIRAP is a robust and efficient coherent population-transfer protocol \cite{10.1063/1.458514, RevModPhys.89.015006}. The Raman processes are typically associated with $\Lambda$-type linkages in atomic systems. The STIRAP is characterized by two counterintuitive pulses, named the pump pulse and the Stokes pulse, which drive the $|3\rangle\leftrightarrow|1\rangle$ and $|3\rangle\leftrightarrow|2\rangle$ transitions of a $\Lambda$ type three-level system.  There are many theoretical and experimental studies on the implementation of STIRAP in SQCS \cite{PhysRevLett.95.087001, SIEWERT2006435, PhysRevB.87.214515, PhysRevB.91.224506, PhysRevA.93.051801, Xu2016, Kumar2016}. This section discusses the implementation of STIRAP in a doubly dressed polariton state-engineered $\Lambda$ system \cite{PhysRevA.110.023716}, which is introduced in Sec. (\ref{DS}). Unlike the EIT and ATS, the pump and the Stokes pulses in STIRAP are chosen to be time-dependent with suitable envelopes. The Hamiltonian for the resonant STIRAP protocol with two Gaussian pulses interacting with the $\Lambda$ system in the ($|1\rangle, |3\rangle, |2\rangle$) basis can be represented by a matrix

\begin{equation}\label{N}
    \hat{H}(t)=\frac{\hbar}{2}
    \begin{pmatrix}
    0 & \Omega_p(t) & 0\\
    \Omega_p(t) & 0 & \Omega_s(t)\\
    0 & \Omega_s(t) & 0
    \end{pmatrix},
\end{equation}

where all the detunings are considered to be zero for simplicity. The instantaneous eigenvalues of this Hamiltonian are
\begin{subequations}\label{O}
\begin{align}
    E_0 & = 0 \quad E_{\pm}=\pm\hbar\Omega_0(t)/2\label{xviii},\\
    \Omega_0(t) & =\sqrt{\Omega^2_p(t)+\Omega^2_s(t)}\label{xix},
\end{align}
\end{subequations}
The instantaneous eigenstates associated with these eigenvalues are 
\begin{equation}\label{P}
    |n_0(t)\rangle=
    \begin{pmatrix}
        \cos{\theta(t)}\\
        0\\
        -\sin{\theta(t)}
    \end{pmatrix},
    |n_{\pm}(t)\rangle=\frac{1}{\sqrt{2}}
    \begin{pmatrix}
        \sin{\theta(t)}\\
        \pm 1\\
        \cos{\theta(t)}
    \end{pmatrix},
\end{equation}
where $\tan{\theta(t)}=\Omega_p(t)/\Omega_s(t)$. The population can be coherently transferred from state $|1\rangle$ to $|2\rangle$ without populating the intermediate state $|3\rangle$ by adiabatically following the dark state $|n_0(t)\rangle$, under the local adiabatic condition $|\dot{\theta}|\ll|\Omega_0|$ \cite{PhysRevA.40.6741}. We can choose the commonly used Gaussian envelopes for the pump and Stokes pulses given by
\begin{subequations}\label{Q}
\begin{align}
    \Omega_p(t)= & \Omega_p e^{-\frac{t^2}{2\sigma^2}}\label{xx},\\
    \Omega_s(t)= & \Omega_s e^{-\frac{(t-t_s)^2}{2\sigma^2}}.\label{xxi}
\end{align}
\end{subequations}
where $\Omega_p$ and $\Omega_s$ are the peak Rabi frequencies of the pump and Stokes pulses, respectively. Both the pulses have equal pulsewidth $\sigma$, and the pump pulse lags the Stokes pulse by a time $t_s$. The time evolution of populations via the STIRAP protocol is discussed along with that of saSTIRAP.

\begin{figure}[t]
    \centering
    \includegraphics[width=\linewidth]{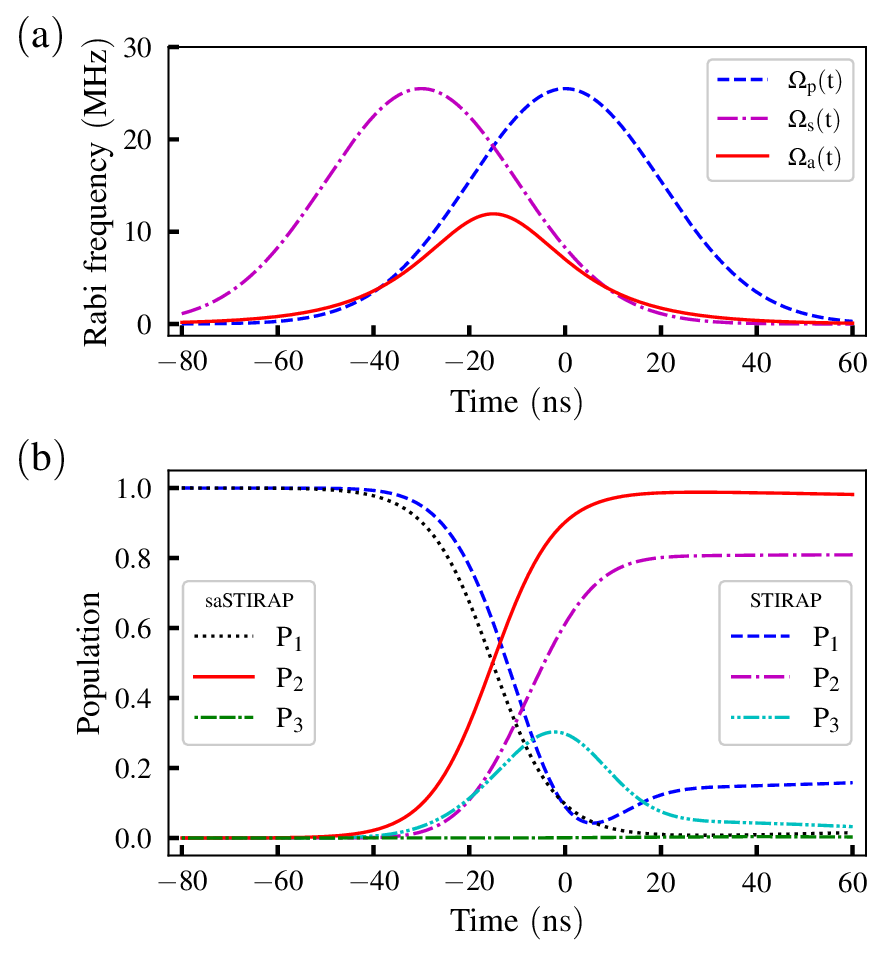}
    \caption{(a) The pulse sequence of the three external driving fields $\Omega_p(t),\Omega_s(t)$, and $\Omega_a(t)$. (b) The time evolution of the populations $P_1,P_2$, and $P_3$ of the states $|1\rangle,|2\rangle$, and $|3\rangle$ during the STIRAP and saSTIRAP protocols, respectively.}
    \label{fig:fig15}
\end{figure}

\subsection{saSTIRAP}
The STIRAP protocol is robust against small fluctuations in experiments and is a powerful technique for coherent population transfer. However, the adiabatic condition renders this protocol more susceptible to decoherence and losses. The recently developed so-called STA techniques aim to tackle this type of challenge \cite{RevModPhys.91.045001}.  The STA methods, such as Berry's counterdiabatic (CD)/transitionless driving \cite{Berry_2009, doi:10.1126/sciadv.aau5999}, the Lewis-Riesenfeld Invariant (LWI) method \cite{10.1063/1.1664991}, and the application of STA by mapping to an effective two-level system \cite{PhysRevA.94.063411}, can be utilized to accelerate the STIRAP protocol. These STA methods enable fast, coherent population transfer, dubbed saSTIRAP, from state $|1\rangle$ to $|2\rangle$, bypassing the strict adiabatic condition for STIRAP. Here, we introduce the counterdiabatic drive 
\begin{equation}\label{R}
    \hat{H}^{CD}(t)=i\hbar\sum\limits_n[|\partial_t n(t)\rangle\langle n(t)|-\langle n(t)|\partial_t n(t)\rangle|n(t)\rangle\langle n(t)|],
\end{equation}
where the dynamic basis states $|n(t)\rangle=(|n_0(t)\rangle,|n_{\pm}(t)\rangle)$, which are introduced in Sec. (\ref{ST}). With a straightforward calculation using the driving pulses given in Eq. (\ref{Q}), one can obtain the final Hamiltonian along with the CD drive
\begin{equation}\label{S}
    \hat{H}^{CD}(t)=\frac{\hbar}{2}
    \begin{pmatrix}
        0 & \Omega_p(t) & i\Omega_a(t)\\
        \Omega_p(t) & 0 & \Omega_s(t)\\
        -i\Omega_a(t) & \Omega_s(t) & 0
    \end{pmatrix},
\end{equation}
where the counterdiabatic drive term is
\begin{equation}\label{T}
    \Omega_a(t)= 2\dot{\theta}=-\frac{t_s}{\sigma^2}\sech{\left[-\frac{t_s}{\sigma^2}{\left(t-\frac{t_s}{2}\right)}\right]}.
\end{equation}
The dipole allowed $|2\rangle\leftrightarrow|1\rangle$ transition in the doubly dressed polariton basis facilitates the implementation of the CD drive in a closed-loop $\Lambda$-system configuration \cite{PhysRevA.110.023716} as shown in Fig. \ref{fig:fig14}. The dynamics of the saSTIRAP system can be studied by solving the master equation (\ref{J}) with the modified Hamiltonian in Eq. (\ref{S}). The pulse sequence of the pump and Stokes fields given in Eq. (\ref{Q}) and the CD drive in Eq. (\ref{T}) are illustrated in Fig. \ref{fig:fig15} (a). The dynamical time-evolution of populations $P_1,P_2$ and $P_3$ of the doubly-dressed polariton states $|1\rangle,|2\rangle$ and $|3\rangle$ are plotted in Fig. \ref{fig:fig15} (b), respectively. The saSTIRAP protocol is faster, more efficient, and robust than STIRAP for coherent population transfer between the two quantum states $|1\rangle$ and $|2\rangle$ without populating the intermediate state $|3\rangle$ \cite{PhysRevA.110.023716, Zheng2022, doi:10.1126/sciadv.aau5999}.  

The STIRAP-inspired quantum gates constitute a broader class of geometric and holonomic quantum gates \cite{ZHANG20231}. The geometric phase and quantum holonomies are global quantities that depend only on the evolution of quantum paths, and the quantum gates based on them are robust against certain kinds of errors. There are many recent works that explore the realization of geometric quantum gates in SQCs \cite{AbdumalikovJr2013, PhysRevApplied.19.034071, PhysRevLett.121.110501, PhysRevLett.122.080501, PhysRevLett.124.230503, PhysRevApplied.20.054047, PhysRevA.109.042615, PhysRevApplied.23.014044, fn6y-byqc}. Moreover, STIRAP-inspired protocols in SQCs have been used for applications including the generation of entangled states \cite{Wu2017}, microwave photon generation \cite{Premaratne2017}, quantum control \cite{Zhou2017}, and quantum state engineering \cite{PhysRevA.111.012623}, etc.

\section{Conclusion}\label{Sec:V}
This review article provides an overview of the remarkable progress toward coherent control in macroscopic SQCs. We began by quantizing the EM field in vacuum. Then, we proceed to discuss the quantization of an LC oscillator, followed by that of a one-dimensional transmission line. 
The Lindblad master equation provides a powerful framework for describing the interaction of quantum systems with their surrounding environment, treating them as open quantum systems OQS. We introduced the basic concepts of superconductivity and the Josephson junction, followed by the basic Josephson junction-based quantum bits, or artificial atoms, which enable the development of novel platforms for observing quantum optical phenomena in macroscopic SQCs. The quantization of lumped-element circuits and transmission lines provides a foundational framework for treating macroscopic devices as artificial circuit QED \cite{RevModPhys.93.025005, 10.1093/acprof:oso/9780199681181.003.0003} and waveguide QED systems \cite{RevModPhys.95.015002}. Dressed states in circuit QED systems are a useful resource for engineering multi-level structures, such as $\Lambda$ systems. Quantum-optical phenomena such as EIT and ATS, which depend on coherent control of quantum coherence, have been implemented in the $\Lambda$ system in the doubly-dressed polariton basis. Furthermore, the implementation of the adiabatic coherent population transfer protocol STIRAP and the accelerated saSTIRAP in the dressed state engineered $\Lambda$ system is discussed. The implementation of these key quantum optical phenomena in SQCs has widespread applications in microwave quantum memory, quantum state preparation, and engineering, among others. However, the SQCs are a versatile platform for demonstrating a wide range of quantum optical phenomena beyond those discussed in this review article \cite{GU20171}. The SQCs can be integrated with other solid-state quantum systems, such as nanomechanics, magnonics, Diamond N-V center spins, and quantum dots, to leverage the strengths of individual systems to explore novel quantum phenomena \cite{RevModPhys.85.623, Clerk2020}. Fully commercial quantum computing \cite{, Wendin_2017}, quantum simulation \cite{PRXQuantum.2.017003}, and quantum metrology \cite{Danilin_2024, RevModPhys.89.035002} remain the ultimate goals as the developments in quantum technologies \cite{4rf7-9tfx}, quantum error correction \cite{RevModPhys.95.045005, RevModPhys.87.307}, and scalable quantum memory networks \cite{doi:10.1126/science.aam9288, https://doi.org/10.1002/lpor.202100219} continue to advance.  

\nocite{*}
\bibliography{review}% Produces the bibliography via BibTeX.

\end{document}